\documentclass{mn2e}

\usepackage{epsfig}

\title[Velocity fields in $\epsilon$~Eri]{A study of velocity fields in the 
transition region of {\boldmath $\epsilon$} Eri (K2 V)}
\author[Sim \& Jordan]{S. A. Sim$^{1,2}$\thanks{s.sim@imperial.ac.uk}, 
C. Jordan$^{1}$\\
$^1$ Department of Physics (Theoretical Physics), University of Oxford, 1 
Keble Road, Oxford, OX1 3NP, UK\\
$^2$ Astrophysics Group, Imperial College London, Blackett Laboratory,
Prince Consort Road, London, SW7 2BW}

\date{2003 Jan 21}

\pagerange{\pageref{firstpage}--\pageref{lastpage}}
\pubyear{2003}

\begin{document}
\maketitle
\label{firstpage}

\begin{abstract}

Analyses of the widths and shifts of optically thin emission lines in 
the ultraviolet spectrum of the active dwarf $\epsilon$~Eri (K2 V) are
presented. The spectra were obtained using the Space Telescope Imaging 
Spectrograph on the {\it Hubble Space 
Telescope} and the {\it Far Ultraviolet Spectroscopic Explorer}.
 The line widths are used to find the 
non-thermal energy density and its variation with temperature from the 
chromosphere to the upper transition region. The energy fluxes that could be
 carried by Alfv\'en and acoustic waves are investigated, to test their 
possible roles in coronal heating. Acoustic waves do not appear to be a viable
means of coronal heating. There is, in principle, ample flux in Alfv\'en waves,
but detailed calculations of wave propagation are required before definite 
conclusions can be drawn about their viability. The high 
sensitivity and spectral resolution of the above instruments have
allowed two-component Gaussian fits to be made to the profiles of 
the stronger transition region lines. The broad and narrow components which 
result share some similarities with those observed in the Sun, but in 
$\epsilon$~Eri the broad component is {\it redshifted} relative to the 
narrow component and contributes more to the total line flux.  The possible
origins of the two components and the energy fluxes implied are discussed.
On balance our results support the conclusion of Wood, Linsky \& Ayres,
that the narrow component is related to Alfv\'en waves reaching to the corona,
but the origin of the broad component is not clear.

\end{abstract}

\begin{keywords} 
stars: individual ($\epsilon$~Eridani) - stars: late-type - stars: 
chromospheres - stars: coronae - line: profiles
\end{keywords}

\renewcommand{\thefootnote}{\fnsymbol{footnote}} 

\section{Introduction}

The observed widths (and profiles) of emission lines, formed in stellar
chromospheres, transition regions and coronae, provide valuable information
on possible heating processes, in addition to that deduced from the emission
line fluxes. Observations of solar line widths in the ultraviolet (uv) part 
of the spectrum have been carried out since 1970 (Berger, Bruner \& Stevens 
1970; Bruner et al. 1970) and much of the basic behaviour of these line widths
was established from work in the 1970s using spectra obtained from rocket 
flights (Boland et al. 1973, 1975; Kjeldseth-Moe \& Nicolas 1977) and from the
Naval Research Laboratory's (NRL) spectrograph on {\it Skylab} 
(Doschek et al. 1976; Feldman et al. 1976). Further work was carried out with
instruments on {\it Orbiting Solar Observatory IV} ({\it OSO-IV}) and the
{\it Solar Maximum Mission} ({\it SMM}). These early observations
showed that line broadening is present in excess of the thermal Doppler width
and that when interpreted in terms of a wave energy flux, this flux is
essentially constant through the transition region from $T_{e} \simeq 10^4$~K
to $\simeq 10^5$~K (where $T_e$ is the electron temperature). The observed 
line widths were found to be very similar in the various solar locations 
studied. Using spectra obtained with the NRL's High
Resolution Telescope and Spectrograph (HRTS), Kjeldseth-Moe \& Nicolas (1977)
found that some emission lines could not be well-fitted with a single 
Gaussian profile, but have excess flux in a broad component. Such
profiles were studied further by Dere \& Mason (1993).
The earliest instruments used photographic 
recording, which has some limitations, but the main results have been
confirmed and extended using the Solar Ultraviolet Measurements of
 Emitted Radiation (SUMER) instrument on the {\it Solar and 
Heliospheric Observatory} ({\it SOHO}). This has allowed line profiles 
to be studied
with a combination of high sensitivity and high spatial and spectral
 resolution, at a variety of locations (see e.g. Doyle et al. 1997; 
Seely et al. 1997; Chae, Sch\"uhle \& Lemaire 1998). Peter (2000a,b;
2001) has 
made systematic measurements of line widths and shifts in both 
supergranulation cell boundary and cell interior regions. We make comparisons 
with the results of Chae et al. (1998) and Peter (2000a,b; 2001) in later 
sections of this paper.

Similar studies of line widths and line shifts were made for a few late-type 
stars observable at high resolution with the {\it International Ultraviolet 
Explorer} ({\it IUE}) (see e.g. Ayres et al. 1983a,b; Brown et
al. 1984; Jordan et al. 1987). Significant 
improvements over these were obtained using the Goddard High Resolution
Spectrograph (GHRS) on the {\it Hubble Space Telescope} ({\it HST}),
and, most 
recently, using the greater spectral coverage and sensitivity of the 
Space Telescope Imaging Spectrograph (STIS). The observations with {\it IUE}
 showed the presence of line broadening in G/K-type dwarfs which is similar 
to that in the Sun, and broader line widths in evolved stars with hot 
coronae. Observations of dwarf and giant stars with the GHRS have
shown the presence of excess flux in 
line wings, compared with single Gaussian fits to the profiles (Linsky
\& Wood 1994; Wood, Linsky \& Ayres 1997, and references therein); the 
contribution of this broad component appears to increase with indicators of
 stellar activity. 

The causes of the observed line broadening are not known, but several
possible processes have been proposed. The excess width
of the lines, compared with the thermal broadening, is commonly attributed to
turbulence, and observed line profiles have been used to find the most-probable
turbulent velocities or the root-mean-square (r.m.s.) velocities. These have 
been used to find non-thermal energy densities and the energy flux carried by
the non-thermal motions, assuming that they are associated with the passage 
of waves through the atmosphere. (See e.g. Brown et al. 1984; Jordan et al.
 1987; Wood et al. 1997 for stellar applications). When
 non-Gaussian profiles have been observed, the excess flux in the wings has
 been attributed to a broad component associated with `explosive
events' (`microflaring') (Linsky \& Wood 1994), or, in the Sun, to the 
passage of Alfv\'en waves (Peter 2001). 
The effect of acoustic waves on line profiles has also been discussed 
(Bruner \& McWhirter 1979).

The possible origins of waves in the solar atmosphere and the processes by 
which these and other mechanisms, in particular magnetic reconnection, may 
cause heating of the solar chromosphere and corona have been the topic of 
many papers and reviews. We have made use of reviews by Browning (1991), 
Narain \& Ulmschneider (1990, 1996) and Zirker (1993).
 
Here we report observations of emission line profiles, widths and line shifts 
in the active dwarf $\epsilon$~Eri (K2~V). These have been made with
the STIS and with the instruments on the {\it Far Ultraviolet Spectroscopic  
Explorer} ({\it FUSE}). The data presented here are the most complete so
far published for a main-sequence star, other than the Sun.   

The new observations and data reduction are described in Section 2. Analyses
 in terms of single Gaussian fits to the
line profiles and the implications of the line widths for the energy
fluxes carried by Alfv\'en waves and acoustic waves are discussed 
in Section 3.
The results of the measurements of line shifts are given in Section 4. 
Some of the line profiles are fitted better by 2-component profiles
rather than single Gaussian fits; the widths and shifts derived for the
two components and their possible interpretations are described in Section 5.
Our conclusions are summarized in Section 6.

\section{Observations and data reduction}

The STIS observations of $\epsilon$ Eri have been described by Jordan et 
al. (2001a). The observations made with {\it FUSE} were obtained from
the archive. In using the latter, we restrict our analyses to the lines of 
C\,{\sc ii} and O\,{\sc vi}, since these were obtained with the 
1a lithium-fluoride detector, 
which is the one used to obtain the pointing on the star. 
The emission lines studied and the transitions involved 
 are listed in Table 1, together with their rest wavelengths ($\lambda_{0}$),
 observed wavelengths ($\lambda_{\mbox{\scriptsize obs}}$) and line 
widths ($\Delta \lambda$), which are the full-width at half-maximum (FWHM). 
The rest wavelengths are taken from Kurucz and Bell (1995) or from
Griesmann \& Kling (2000), with the
exception of that for the Fe~{\sc xii} line, which is based on the solar
 wavelength (see Jordan et
al. 2001a). The observed wavelengths and comparisons with the
rest values given by the above authors and by Kelly (1987) are
discussed in Section 4. 

The widths given in Table 1 were measured by 
making single-Gaussian fits to the observed spectra. Except at wavelengths 
above $\simeq 1800$~\AA, the background level is close to zero. At $\lambda
\ge 1800$~\AA, the background was determined from the nearby continuum. 
Likely errors (based on the wavelength resolution of the appropriate 
echelle grating, the uncertainty in the background level, the goodness of the 
fit and comparison with fluxes found by direct integration) are given 
for each measurement. For some of the weakest lines the signal-to-noise 
ratio was judged to be too small to give reliable line widths. 
In addition to the Gaussian fits, for the stronger lines, interactive 
measurements of the width at half the peak amplitude were made. For 
most of the lines these agree with 
the Gaussian fit values, within the error bars, but for the resonance 
lines of C\,{\sc iv}, N\,{\sc v}, O\,{\sc vi} and Si\,{\sc iv} they are 
significantly smaller indicating that these profiles are not adequately
described by a single Gaussian fit. 
These half peak-amplitude widths are also given in Table 1. 
It is found that the profiles of these lines are described much better by 
two-Gaussian fits (see Section 5). 

The STIS instrumental widths are taken from the STIS WWW site, and are 1.1
-- 2 pixels, depending on the wavelength and grating used. For the
E140M grating, one pixel is typically $\sim 0.015$~\AA, and this
instrumental broadening does not contribute significantly to the
observed line widths. For the E230M grating, one pixel is typically
$\sim 0.03$~\AA, which leads to significant contributions to the
widths of longer wavelength lines observed with this grating (see Table 1 in
Jordan et al. 2001b). The {\it FUSE} observations have a resolution
$R = 20,000$ (Blair \& Andersson 2001), corresponding to $\Delta
\lambda \sim 0.09$~\AA.

The effect of rotational broadening has been considered but is judged
to be negligible. For $\epsilon$ Eri, $v \sin i < 2$~km~s$^{-1}$ (Saar
\& Osten 1997) which is too small (compared to the instrumental
resolution) to be detectable; it is noted however, that only a factor
of a few increase in the spectral resolution would make this detectable.

\begin{table*}
\caption{Transitions, wavelengths and widths of the emission lines studied.}
\begin{tabular}{l c l l l} \\ \hline
Ion & Transition & $\lambda_{0}$ (\AA) & $\lambda_{\mbox{\scriptsize
obs}}$ (\AA)& $\Delta \lambda_{\mbox{\scriptsize FWHM}}$ (\AA)\\ \hline
C {\sc i} & 2s$^{2}$2p$^{2}$ $^{3}$P$_{0}$ -- 2s2p$^{3}$ $^{3}$D$_{1}$ &
1560.310 & $1560.331 \pm 0.008$ & $0.135 \pm 0.008$$^b$ \\
C {\sc i} & 2p$^{2}$ $^{3}$P$_{1}$ -- 2p3s $^{3}$P$_{2}$ &
1656.267 & $1656.273 \pm 0.009$ & $0.191 \pm 0.008$$^{b}$ \\
C {\sc i}   & 2p$^{2}$ $^{1}$D$_{2}$ -- 2p3s $^{3}$P$_{1}$  &  
1993.620 & $1993.642 \pm 0.016$ & $0.110 \pm 0.016$\\
C {\sc iii} & 2s2p $^3$P$_1$ -- 2p$^2$ $^3$P$_2$  &
1174.933 & $1174.949 \pm 0.006$ & $0.157 \pm 0.015$ \\
C {\sc iii} & 2s2p $^3$P$_0$ -- 2p$^2$ $^3$P$_1$  &
1175.263 & $1175.281 \pm 0.006$ & $0.155 \pm 0.015$ \\
C {\sc iii} & 2s2p $^3$P$_1$ -- 2p$^2$ $^3$P$_0$  &
1175.987 & $1176.002 \pm 0.006$ & --$^{c}$ \\
C {\sc iii} & 2s2p $^3$P$_2$ -- 2p$^2$ $^3$P$_1$  &
1176.370 & $1176.384 \pm 0.006$ & --$^{c}$ \\
C {\sc iv}  & 2s $^{2}$S$_{1/2}$ -- 2p $^{2}$P$_{3/2}$   &
1548.204 & $1548.224 \pm 0.008$ & $0.207 \pm 0.008$\\
&&&& $0.183 \pm 0.008$$^{a}$\\
C {\sc iv}  & 2s $^{2}$S$_{1/2}$ -- 2p $^{2}$P$_{1/2}$   &
1550.781 & $1550.788 \pm 0.008$ & $0.210 \pm 0.008$\\
&&&& $0.177 \pm 0.008$$^{a}$\\
N {\sc iv}  & 2s$^{2}$ $^{1}$S$_{0}$ -- 2s2p $^{3}$P$_{1}$   &
1486.496 & $1486.521 \pm 0.017$ & $0.210 \pm 0.050$\\
N {\sc v}  & 2s $^{2}$S$_{1/2}$ -- 2p $^{2}$P$_{3/2}$   &
1238.821& $1238.828 \pm 0.007$ & $0.178 \pm 0.007$\\
&&&& $0.155 \pm 0.007$$^{a}$ \\
N {\sc v}  & 2s $^{2}$S$_{1/2}$ -- 2p $^{2}$P$_{1/2}$   &
1242.804& $1242.822 \pm 0.007$ & $0.172 \pm 0.007$\\
&&&& $0.149 \pm 0.007$$^{a}$ \\
O {\sc i}  & 2p$^{4}$ $^{3}$P$_{2}$ -- 2p$^{3}$($^{4}$S)3s $^{5}$S$_{2}$ &
1355.598 & $1355.607 \pm 0.007$ & $0.065 \pm 0.007$\\
O {\sc i}  & 2p$^{4}$ $^{3}$P$_{1}$ -- 2p$^{3}$($^{4}$S)3s $^{5}$S$_{2}$ &
1358.512 & $1358.523 \pm 0.007$ & $0.056 \pm 0.010$\\
O {\sc iii} & 2s$^{2}$2p$^{2}$ $^{3}$P$_{1}$ -- 2s2p$^{3}$ $^{5}$S$_{2}$ & 
1660.809 & $1660.847 \pm 0.018$ & --$^{c}$\\
O {\sc iii} & 2s$^{2}$2p$^{2}$ $^{3}$P$_{2}$ -- 2s2p$^{3}$ $^{5}$S$_{2}$ & 
1666.150 & $1666.169 \pm 0.018$ & $0.217 \pm 0.030$\\
O {\sc iv} & 2s$^{2}$2p $^{2}$P$_{1/2}$ -- 2s2p$^{2}$ $^{4}$P$_{1/2}$ &
1399.780& $1399.797 \pm 0.015$ & --$^{c}$\\
O {\sc iv} & 2s$^{2}$2p $^{2}$P$_{3/2}$ -- 2s2p$^{2}$ $^{4}$P$_{5/2}$ &
1401.157 & $1401.186 \pm 0.008$ & $0.191 \pm 0.015$ \\
O {\sc iv} & 2s$^{2}$2p $^{2}$P$_{3/2}$ -- 2s2p$^{2}$ $^{4}$P$_{1/2}$ &
1407.382 & $1407.429 \pm 0.015$ & --$^{c}$\\
O {\sc v} & 2s2p $^{1}$P$_{1}$ -- 2p$^{2}$ $^{1}$D$_{2}$ &
1371.296 & $1371.327 \pm 0.007$ & $0.193 \pm 0.011$ \\
O {\sc vi} & 2s $^{2}$S$_{1/2}$ -- 2p$^{2}$P$_{3/2}$ & 
1031.912  & see text   & $0.194 \pm 0.015$   \\
&&&& $0.180 \pm 0.015$$^a$ \\
O {\sc vi} & 2s $^{2}$S$_{1/2}$ -- 2p$^{2}$P$_{1/2}$ &
1037.613  & see text   & $0.185 \pm 0.015$   \\
&&&& $0.165 \pm 0.015$$^a$ \\
Al {\sc ii} & 3s$^{2}$ $^{1}$S$_{0}$ -- 3s3p $^{1}$P$_{1}$ &
1670.789 & $1670.802 \pm 0.009$ & $0.213 \pm 0.018$$^{b?}$\\
Si {\sc ii} & 3s$^{2}$3p $^{2}$P$_{3/2}$ -- 3s3p$^{2}$ $^{2}$P$_{3/2}$ &
1194.500 & $1194.499 \pm 0.013$ & $0.188 \pm 0.040$$^{b}$ \\
Si {\sc ii} & 3s$^{2}$3p $^{2}$P$_{3/2}$ -- 3s3p$^{2}$ $^{2}$P$_{1/2}$ &
1197.394 & $1197.389 \pm 0.013$ & --$^{c}$\\
Si {\sc ii} & 3p $^{2}$P$_{3/2}$ -- 3d $^{2}$D$_{5/2}$ &
1264.738 & $1264.747 \pm 0.007$ & $0.184 \pm 0.007$$^{b}$ \\
Si {\sc ii} & 3p $^{2}$P$_{3/2}$ -- 3d $^{2}$D$_{3/2}$ &
1265.002 & $1265.012 \pm 0.007$ & $0.157 \pm 0.007$$^{b}$ \\
Si {\sc ii} & 3p $^{2}$P$_{1/2}$ -- 4s $^{2}$S$_{1/2}$ &
1526.707 & $1526.717 \pm 0.008$ & $0.173 \pm 0.008$$^{b}$ \\
Si {\sc ii} & 3p $^{2}$P$_{3/2}$ -- 4s $^{2}$S$_{1/2}$ &
1533.431 & $1533.435 \pm 0.008$ & $0.181 \pm 0.008$$^{b}$ \\
Si {\sc ii} & 3s$^{2}$3p $^{2}$P$_{1/2}$ -- 3s3p$^{2}$ $^{2}$D$_{3/2}$ &
1808.013 & $1808.069 \pm 0.015$ & $0.136 \pm 0.015$$^{b?}$\\
Si {\sc ii} & 3s$^{2}$3p $^{2}$P$_{3/2}$ -- 3s3p$^{2}$ $^{2}$D$_{5/2}$ &
1816.928 & $1816.970 \pm 0.015$ & $0.142 \pm 0.015$$^{b?}$\\
Si {\sc ii} & 3s$^{2}$3p $^{2}$P$_{3/2}$ -- 3s3p$^{2}$ $^{2}$D$_{3/2}$ &
1817.451 & $1817.491 \pm 0.015$ & $0.117 \pm 0.015$$^{b?}$\\
Si {\sc ii} & 3s$^{2}$3p $^{2}$P$_{1/2}$ -- 3s3p$^{2}$ $^{4}$P$_{1/2}$ &
2335.123 & $2335.109 \pm 0.019$ & $0.129 \pm 0.019$ \\
Si {\sc ii} & 3s$^{2}$3p $^{2}$P$_{3/2}$ -- 3s3p$^{2}$ $^{4}$P$_{1/2}$ &
2350.891 & $2350.896 \pm 0.019$ & $0.124 \pm 0.019$ \\
Si {\sc iii} & 3s3p $^{3}$P$_{1}$ -- 3p$^{2}$ $^{3}$P$_{2}$ &
1294.545 & $1294.578 \pm 0.014$ & -- $^{c}$\\
Si {\sc iii} & 3s3p $^{3}$P$_{0}$ -- 3p$^{2}$ $^{3}$P$_{1}$ &
1296.726 & $1296.758 \pm 0.014$ & -- $^{c}$\\
Si {\sc iii} & 3s3p $^{3}$P$_{1}$ -- 3p$^{2}$ $^{3}$P$_{0}$ &
1301.149 & $1301.168 \pm 0.028$ & -- $^{c}$\\
Si {\sc iii} & 3s3p $^{3}$P$_{2}$ -- 3p$^{2}$ $^{3}$P$_{1}$ &
1303.323 & $1303.349 \pm 0.014$ & -- $^{c}$\\
Si {\sc iii} & 3s$^{2}$ $^{1}$S$_{0}$ -- 3s3p $^{3}$P$_{1}$ &
1892.030 & $1892.106 \pm 0.015$ & $0.194 \pm 0.015$ \\
Si {\sc iv} & 3s $^{2}$S$_{1/2}$ -- 3p $^{2}$P$_{3/2}$   &
1393.760 & $1393.781 \pm 0.008$ & $0.176 \pm 0.008$\\
&&&& $0.143 \pm 0.008$$^{a}$ \\
Si {\sc iv} & 3s $^{2}$S$_{1/2}$ -- 3p $^{2}$P$_{1/2}$   &
1402.773 & $1402.795 \pm 0.008$ & $0.175 \pm 0.008$\\
&&&& $0.155 \pm 0.008$$^{a}$ \\
S {\sc i} & 3p$^{4}$ $^3$P$_{2}$ -- 3p$^{3}$($^4$S)4s $^5$S$_2$ &
1900.287 & $1900.322 \pm 0.016$ & $0.104 \pm 0.016$\\
S {\sc i} & 3p$^{4}$ $^3$P$_{1}$ -- 3p$^{3}$($^4$S)4s $^5$S$_2$ &
1914.697 & $1914.757 \pm 0.016$ & $0.106 \pm 0.016$\\
S {\sc ii} & 3s$^{2}$3p$^{3}$ $^{4}$S$_{3/2}$ -- 3s3p$^{4}$ $^{4}$P$_{3/2}$ &
1253.811 & $1253.820 \pm 0.015$ & $0.108 \pm 0.015$\\
S {\sc ii} & 3s$^{2}$3p$^{3}$ $^{4}$S$_{3/2}$ -- 3s3p$^{4}$ $^{4}$P$_{5/2}$ &
1259.519 & $1259.533 \pm 0.015$ & $0.122 \pm 0.015$\\
Cl {\sc i} & 3p$^5$ $^2$P$_{1/2}$ -- 3p$^{4}$($^3$P)4s $^2$P$_{1/2}$ &
1351.656 & $1351.664 \pm 0.007$ & $0.061 \pm 0.007$\\
Fe {\sc ii} & 3d$^6$(a$^{5}$D)4s a$^{6}$D$_{9/2}$ -- 3d$^5$(a$^{6}$S)4s4p($^3$P) z$^{8}$P$_{9/2}$&
1888.010 & $1888.032 \pm 0.015$ & --$^{c}$\\
Fe {\sc ii} & 3d$^6$(a$^{5}$D)4s a$^{6}$D$_{9/2}$ -- 3d$^5$(a$^{6}$S)4s4p($^3$P) z$^{8}$P$_{7/2}$& 
1901.773 & $1901.799 \pm 0.016$ & $0.130 \pm 0.016$\\
Fe {\sc ii} & 3d$^6$(a$^{5}$D)4s a$^{6}$D$_{7/2}$ -- 3d$^5$(a$^{6}$S)4s4p($^3$P) z$^{8}$P$_{7/2}$&
1915.792 & $1915.817 \pm 0.016$ & --$^{c}$\\
Fe {\sc ii} & 3d$^6$(a$^{5}$D)4s a$^{6}$D$_{7/2}$ -- 3d$^5$(a$^{6}$S)4s4p($^3$P) z$^{8}$P$_{5/2}$& 
1926.240 & $1926.278 \pm 0.016$ & $0.127 \pm 0.016$\\
Fe {\sc ii} & 3d$^6$(a$^{5}$D)4s a$^{6}$D$_{5/2}$ -- 3d$^5$(a$^{6}$S)4s4p($^3$P) z$^{8}$P$_{5/2}$&
1936.794 & $1936.810 \pm 0.016$ & $0.124 \pm 0.016$\\
Fe {\sc ii} & 3d$^6$(a$^{5}$D)4s a$^{6}$D$_{3/2}$ -- 3d$^5$(a$^{6}$S)4s4p($^3$P) z$^{8}$P$_{5/2}$&
1944.134 & $1944.156 \pm 0.016$ & --$^{c}$\\
Fe {\sc xii} & 3p$^{3}$ $^4$S$_{3/2}$ -- 3p$^3$ $^2$P$_{3/2}$ &
1242.00 & $1241.992 \pm 0.007$ & $0.21 \pm 0.04$\\ \hline
\end{tabular}\\
\noindent $^{a}$ Width at half the peak amplitude.
\noindent $^{b}$ Optically thick line.
\noindent $^{c}$ Line too weak to determine the width.\\
\end{table*}

\section{Non-thermal velocities}

Most of the permitted lines, and all of the intersystem lines, formed in the 
transition region are expected to be optically thin. In the transition 
region, line opacities can be estimated from the observed line fluxes and 
the mean electron pressure determined by Jordan et al. (2001b). Intersystem
lines formed in the chromosphere are also estimated to be optically thin.
It is harder to estimate the opacities of permitted chromospheric lines and,
as will be discussed later, some of these are likely to be optically thick,
and will have contributions to their line widths from photon scattering.
Lines to the ground level of atoms and first ions may also be affected
by interstellar absorption. The ratios of lines within multiplets and
the presence of self-reversals have also been used to judge whether or not
lines are optically thin. 

As discussed in Section 1, the widths of optically thin emission lines in 
stellar uv spectra are significantly greater than the thermal Doppler
 width corresponding to the temperature of line formation 
($T_{\mbox{\scriptsize form}}$). When corrected for instrumental broadening
 it is usually assumed that the observed width is the result of thermal and 
non-thermal motions, the latter being referred to as turbulence. 
Since most optically thin line profiles are well described by 
one (or more) Gaussian components, it is assumed that the turbulent 
velocities follow a Maxwellian distribution (and hence produce a Gaussian 
line profile), and the distribution is characterised by its most probable
 velocity $\xi$. This non-thermal contribution to
 the line width is usually the dominant contribution to chromospheric and 
transition region line widths.

The total line width may be expressed as

\begin{equation}
\left( 
{\frac{\Delta \lambda}{\lambda_{0}}} \right)_{\mbox{\scriptsize obs}}^2
=
\left( {\frac{\Delta \lambda}{\lambda_{0}}} 
\right)_{\mbox{\scriptsize therm}}^2+
\left( {\frac{\Delta \lambda}{\lambda_{0}}} 
\right)_{\mbox{\scriptsize inst}}^2
+
\left( {\frac{\Delta \lambda}{\lambda_{0}}} 
\right)_{\mbox{\scriptsize non-them}}^2
\end{equation}
where the subscript ``obs'', ``therm'', ``inst'' and ``non-therm'' refer to 
the observed total width and the contributions from thermal, instrumental and 
non-thermal  broadening respectively. By expressing the thermal width in 
terms of the temperature of line formation and the non-thermal width in terms 
of a most probably velocity $\xi$, equation (1) may be written as

\begin{equation}
\frac{\xi^2}{c^2} = \frac{1}{4 \ln 2} \left[ {\left( {\frac{\Delta
\lambda}{\lambda_{0}}} \right)_{\mbox{\scriptsize obs}}^2 - \left( 
{\frac{\Delta \lambda}{\lambda_{0}}} \right)_{\mbox{\scriptsize
inst}}^2} \right] - \frac{2 k_{\mbox{\scriptsize B}}
T_{\mbox{\scriptsize form}}}{m_{\mbox{\scriptsize H}}
m_{\mbox{\scriptsize ion}} c^{2}}    
\end{equation}
giving
\begin{equation}
\frac{\xi^2}{c^2} = 0.361 \left[ {
\left( {\frac{\Delta \lambda}{\lambda_{0}}} \right)_{\mbox{\scriptsize
 obs}}^2 
 - 
\left( {\frac{\Delta \lambda}{\lambda_{0}}} \right)_{\mbox{\scriptsize inst}}^2
}\right]
- 1.84 \times 10^{-13} \frac{T_{\mbox{\scriptsize form}}}{m_{\mbox{\scriptsize
 ion}}}    
\end{equation}
where $T_{\mbox{\scriptsize form}}$ is in Kelvin,
$m_{\mbox{\scriptsize H}}$ is the mass of hydrogen  and 
$m_{\mbox{\scriptsize ion}}$ is the mass of the emitting ion, relative to 
that of hydrogen.

For the chromospheric lines, formed at low temperatures, 
 the value of $\xi$ is relatively insensitive to the adopted value of 
$T_{\mbox{\scriptsize form}}$, but because these lines are formed over
 a wide range of temperatures, the actual values of
 $T_{\mbox{\scriptsize form}}$ are less meaningful.
For the transition region lines the value of $\xi$ is sensitive
to $T_{\mbox{\scriptsize form}}$, which should be regarded as only an
average value.
The adopted values of $T_{\mbox{\scriptsize form}}$ 
are given in Table 2. For the transition region lines formed above
$\log T_{e} > 4.3$,  the values of $T_{\mbox{\scriptsize form}}$ 
are based on the line contribution functions and the observed emission 
measure distribution (Jordan, Sim \& McMurry 2001c; Sim 2002).
New calculations of the relative ion populations by Sim
(2002) have been used, which take into account the density sensitivity
of di-electronic recombination. This reduces the temperature of
optimum line formation compared with the zero density calculations of
Arnaud \& Rothenflug (1985). Overall, the values of
$T_{\mbox{\scriptsize form}}$ are expected to be accurate to within
about $\pm 0.1$~dex. 
Values for the chromospheric lines of O~{\sc i}, C~{\sc i} and 
Si~{\sc ii} are based on Sim's (2002) calculations with the MULTI 
(Carlsson 1986, Scharmer \& Carlsson 1985) radiative transfer code;
those for Cl~{\sc i}, Al~{\sc ii} and Fe~{\sc ii} are only rough estimates;
the value adopted for Fe~{\sc xii} is that used by Jordan et al. (2001a). 

\begin{table}
\caption{Values of $T_{\mbox{\scriptsize form}}$ adopted.} 
\begin{tabular}{lcc}\hline
Ion & $\lambda_{0}$(\AA) & $\log T_{\mbox{\scriptsize form}}$(K)\\ \hline
C {\sc i} & 1993.62 & 3.7\\
C {\sc iii} & 1170 mult. & 4.7 \\
C {\sc iv} & 1550 mult. & 4.9 \\
N {\sc iv} & 1486.50 & 5.0 \\
N {\sc v} & 1240 mult. & 5.2\\
O {\sc i} & 1356 mult. & 3.8\\
O {\sc iii} & 1666.15 & 4.9\\
O {\sc iv} & 1401.16 & 5.2\\
O {\sc v} & 1371.30 & 5.4\\
O {\sc vi} & 1034 mult. & 5.45 \\
Al {\sc ii} & 1670.79 & 4.2\\
Si {\sc ii} & 1815 mult. & 3.85\\
Si {\sc ii} & 2340 mult. & 3.65\\
Si {\sc iii} & 1892.03 & 4.45 \\
Si {\sc iv} & 1400 mult. & 4.8 \\
S {\sc i} & 1907 mult. & 3.7 \\
S {\sc ii} & 1255 mult. & 4.2 \\
Cl {\sc i} & 1351.66& 3.7 \\
Fe {\sc ii} & 1915 mult. & 3.85\\
Fe {\sc xii} & 1242.00 & 6.15 \\ \hline
\end{tabular}
\end{table}

The values of $\xi$ found using equation (3) are given in Table 3.
The errors on $\xi$ include both the measurement errors
from Table 1 and an uncertainty in 
$\log T_{\mbox{\scriptsize form}}$ of $\pm 0.1$~dex. The value in 
parenthesis after the total error is the contribution  
from the uncertainty in $\log T_{\mbox{\scriptsize form}}$ alone. 
The observed line positions have been converted to velocity shifts
using the rest wavelengths $\lambda_{0}$ given in Table 1.

\begin{table*}
\caption{Derived line shifts ($v$) and turbulent velocities ($\xi$).
 The error estimates for $\xi$ include the total uncertainty due to
the measurement errors and an assumed uncertainty in 
$\log T_{\mbox{\scriptsize form}}$ of $\pm 0.1$~dex. The errors in 
parentheses indicate the uncertainty due to 
$\Delta \log T_{\mbox{\scriptsize form}}$ alone.}
\begin{tabular}{llll}\hline
Ion & $\lambda_{0}$ (\AA) & $v$ (km s$^{-1}$) & $\xi$ (km s$^{-1}$) \\ \hline
C {\sc i} & 
1560.310 & $+4.0 \pm 1.5$&-- \\
C {\sc i} &
1656.267 & $+1.1 \pm 1.6$ &--\\
C {\sc i}   & 
1993.620 & $+3.3 \pm 2.4$ & 7.55$^{+1.90(0.10)}_{-2.20(0.12)}$\\[2mm]
C {\sc iii} & 
1174.933 & $+4.1 \pm 1.5$ & 19.50$^{+2.94(0.45)}_{-3.29(0.59)}$\\[2mm]
C {\sc iii} & 
1175.263 & $+4.6 \pm 1.5$ & 22.11$^{+2.74(0.32)}_{-2.96(0.41)}$\\
C {\sc iii} & 
1175.987 & $+3.8 \pm 1.5$ &--\\
C {\sc iii} & 
1176.370 & $+3.6 \pm 1.5$&--\\
C {\sc iv}  & 
1548.204 & $+3.9 \pm 1.6$ & 21.57$^{+1.53(0.52)}_{-1.75(0.67)}$\\[2mm]
C {\sc iv}  & 
1550.781 & $+1.4 \pm 1.5$ & 21.91$^{+1.52(0.51)}_{-1.73(0.66)}$\\[2mm]
N {\sc iv}  & 
1486.496 & $+5.0 \pm 3.4$ & 22.89$^{+7.00(0.52)}_{-8.01(0.68)}$\\[2mm]
N {\sc v}  & 
1238.821& $+1.7 \pm 1.7$ & 21.82$^{+2.02(0.86)}_{-2.43(1.13)}$\\[2mm]
N {\sc v}  & 
1242.804& $+4.3 \pm 1.7$ & 20.68$^{+2.08(0.91)}_{-2.52(1.21)}$\\[2mm]
O {\sc i}  &
1355.598 & $+2.0 \pm 1.5$ & 7.79$^{+1.10(0.09)}_{-1.17(0.11)}$\\[2mm]
O {\sc i}  &
1358.512 & $+2.4 \pm 1.5$ & 6.42$^{+1.58(0.10)}_{-1.77(0.13)}$\\
O {\sc iii} & 
1660.809 & $+6.9 \pm 3.3$&--\\
O {\sc iii} & 
1666.150 & $+3.4 \pm 3.3$ & 21.55$^{+3.83(0.39)}_{-4.19(0.50)}$\\
O {\sc iv} & 
1399.780& $+3.6 \pm 3.2$&--\\
O {\sc iv} & 
1401.157 & $+6.2 \pm 1.7$ & 20.82$^{+2.96(0.79)}_{-3.50(1.04)}$\\
O {\sc iv} & 
1407.382 & $+10. \pm 3.2$&--\\
O {\sc v} & 
1371.296 & $+6.8 \pm 1.5$ & 19.44$^{+3.07(1.33)}_{-3.97(1.81)}$\\[2mm]
O {\sc vi} & 
1031.912   &  see text  & 24.27$^{+4.60(1.20)}_{-5.75(1.61)}$\\[2mm]
O {\sc vi} &
1037.613   &  see text  & 22.03$^{+4.83(1.32)}_{-6.30(1.78)}$\\[2mm]
Al {\sc ii} & 
1670.789 & $+2.3 \pm 1.6$ & 22.67$^{+2.01(0.04)}_{-2.03(0.06)}$\\[2mm]
Si {\sc ii} &
1194.500 & $-0.3 \pm 3.3$&--\\
Si {\sc ii} &
1197.394 & $-1.3 \pm 3.3$&--\\
Si {\sc ii} &
1264.738 & $+2.1 \pm 1.7$&--\\
Si {\sc ii} &
1265.002 & $+2.4 \pm 1.7$&--\\ \hline
\end{tabular}\hspace{1.cm}
\begin{tabular}{llll}\hline
Ion & $\lambda_{0}$ (\AA) & $v$ (km s$^{-1}$) & $\xi$ (km s$^{-1}$) \\ \hline
Si {\sc ii} &
1526.707 & $+2.0 \pm 1.6$&--\\
Si {\sc ii} &
1533.431 & $+0.8 \pm 1.6$&--\\
Si {\sc ii} &
1808.013 & $+9.3 \pm 2.5$ & 12.01$^{+1.70(0.04)}_{-1.77(0.05)}$\\[2mm]
Si {\sc ii} &
1816.928 & $+6.9 \pm 2.5$& 12.62$^{+1.67(0.03)}_{-1.74(0.04)}$\\[2mm]
Si {\sc ii} &
1817.451 & $+6.6 \pm 2.5$& 9.78$^{+1.76(0.04)}_{-1.89(0.06)}$\\[2mm]
Si {\sc ii} & 
2335.123 & $-1.8 \pm 2.4$& 8.07$^{+1.78(0.03)}_{-1.96(0.04)}$\\[2mm]
Si {\sc ii} & 
2350.891 & $+0.6 \pm 2.4$& 7.53$^{+1.80(0.04)}_{-2.01(0.05)}$\\
Si {\sc iii} & 
1294.545 & $+7.6 \pm 3.2$&--\\
Si {\sc iii} & 
1296.726 & $+7.4 \pm 3.2$&--\\
Si {\sc iii} & 
1301.149 & $+4.4 \pm 6.5$&--\\
Si {\sc iii} & 
1303.323 & $+6.0 \pm 3.2$&--\\
Si {\sc iii} &
1892.030 & $+12. \pm 2.4$& 17.04$^{+1.63(0.10)}_{-1.70(0.13)}$\\[2mm]
Si {\sc iv} & 
1393.760 & $+4.5 \pm 1.7$& 21.78$^{+1.25(0.18)}_{-1.32(0.22)}$\\[2mm]
Si {\sc iv} & 
1402.773 & $+4.7 \pm 1.7$& 21.49$^{+1.24(0.18)}_{-1.32(0.23)}$\\[2mm]
S {\sc i} & 
1900.287 & $+5.5 \pm 2.5$& 7.73$^{+1.89(0.03)}_{-2.12(0.04)}$\\[2mm]
S {\sc i} & 
1914.697 & $+9.4 \pm 2.5$& 7.91$^{+1.86(0.03)}_{-2.07(0.04)}$\\[2mm]
S {\sc ii} &
1253.811 & $+2.2 \pm 3.6$& 15.01$^{+2.27(0.06)}_{-2.33(0.07)}$\\[2mm]
S {\sc ii} &
1259.519 & $+3.3 \pm 3.6$& 17.01$^{+2.24(0.05)}_{-2.28(0.06)}$\\[2mm]
Cl {\sc i} &
1351.656 & $+1.8 \pm 1.6$& 7.50$^{+1.03(0.03)}_{-1.07(0.04)}$\\
Fe {\sc ii}& 
1888.010 & $+3.5 \pm 2.4$&--\\
Fe {\sc ii} &
1901.773 & $+4.1 \pm 2.5$& 10.71$^{+1.73(0.02)}_{-1.81(0.03)}$\\
Fe {\sc ii} &
1915.792 & $+3.9 \pm 2.5$&--\\
Fe {\sc ii} &
1926.240 & $+5.9 \pm 2.5$& 10.21$^{+1.73(0.02)}_{-1.82(0.03)}$\\[2mm]
Fe {\sc ii} &
1936.794 & $+2.5 \pm 2.5$& 9.81$^{+1.73(0.02)}_{-1.84(0.03)}$\\
Fe {\sc ii} &
1944.134 & $+3.4 \pm 2.5$&--\\
Fe {\sc xii} &
1242.00 & $-1.9 \pm 1.7$$^{a}$& 22.46$^{+8.81(1.83)}_{-13.71(2.55)}$\\[3mm] \hline
\end{tabular}

\noindent ~\hspace{4cm}$^{a}$ $\lambda_{0}$ is a solar value. \\
\end{table*}

The values of $\xi$ given in Table 3 are plotted against 
$T_{\mbox{\scriptsize form}}$ in Fig. 1, including only those lines
which are estimated to be optically thin.  Qualitatively, this 
plot is similar to those found for the Sun (see e.g. Jordan 1991; 
Chae et al. 1998); $\xi$ is smallest in the 
chromosphere, rises in the low transition region and then flattens 
off. Comparing
 optically thin lines in common with those of Chae et al. 
(1998), the range of $\xi$ is slightly smaller in $\epsilon$~Eri, the widths 
of the mid-transition region lines are {\it smaller} in $\epsilon$~Eri, but
the widths of the neutral lines are on average {\it larger} in $\epsilon$~Eri.
All the lowest temperature 
chromospheric lines (those of the neutrals) are consistent with 
$\xi \approx 7.5$~km~s$^{-1}$. The widths of the lowest 
temperature lines of Si {\sc ii} (the intersystem multiplet around 
2340~\AA) agree with this mean value.  {\it All} the 
lines with $\log T_{\mbox{\scriptsize form}} > 4.5$ have widths that are  
consistent with a {\it constant} value of $\xi \approx 21.3$~km s$^{-1}$ 
throughout the mid-transition region. The other lines 
indicate a smooth transition between the two regimes. 

There are some differences in the value of $\xi$ for lines formed 
at similar temperatures, which may be due to opacity effects. 
The value of $\xi$ for the Al~{\sc ii} resonance line is larger than 
those for the S~{\sc ii} lines. It is difficult to judge whether or not 
the observed Al~{\sc ii} profile is self-reversed, or contains  
interstellar absorption, but the opacity estimate suggests that small
optical depth effects could occur.
The lines in the Si~{\sc ii} multiplet around 1815~\AA~do not have 
ratios which agree with their optically thin values, and the two lines 
with the largest oscillator strengths (1808.01~\AA~and 1816.93~\AA) are 
slightly broader than the line at 1817.45~\AA~and the intersystem lines of  
Fe~{\sc ii}. Although these Si~{\sc ii} lines are not self-reversed, small
opacity effects could be occurring. For these reasons
these lines of Si~{\sc ii} and Al~{\sc ii} are not used in
subsequent analyses using non-thermal widths.  
Lines from the other multiplets of Si~{\sc ii} have significantly 
larger oscillator strengths than those in the
1815-\AA~multiplet, and should therefore show opacity effects. The
lines at 1533.43~\AA~and 1526.71~\AA~have a flux ratio which is $\simeq
1$, instead of the optically thin value of $2$. The multiplets at
shorter wavelengths have larger or similar oscillator strengths (as
given in CHIANTI, Landi et al. 1999), and are therefore also expected
to be optically thick. The widths of these other lines of Si~{\sc ii}
are included in Table 1, but are not included in Fig. 1.

\begin{figure}
\epsfig{file=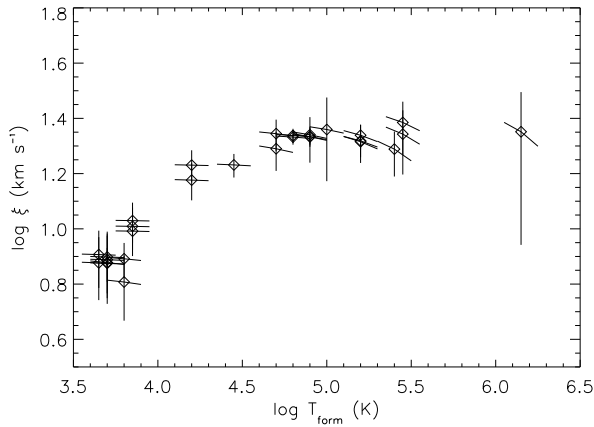, width=8.6cm}
\caption{$\log \xi$ versus $\log T_{\mbox{\scriptsize form}}$. The vertical error bars indicate the total uncertainty in the deduced values of $\xi$ (including measurement errors and uncertainty in $T_{\mbox{\scriptsize form}}$) and the sloping bars indicate the effects of uncertainties in $T_{\mbox{\scriptsize form}}$.}
\end{figure}

\subsection{Kinetic energy density}

The interpretation of the line broadening in spatially integrated
stellar spectra is subject to some uncertainties. Solar observations
can be used to investigate the behaviour of line widths at different
positions on the Sun. In both early studies (see references in Section
1) and in more recent work with SUMER (see e.g. Peter 1999), when
single Gaussian fits are made to the line profiles there is
little systematic difference between the 
widths at different solar locations and
distances from the limb. In this respect, using spatially averaged
spectra should give a good estimate of the typical line widths. These
solar observations also show that there are no preferred directions of
the motions contributing to the widths. However, in the Sun, there are
line (Doppler) shifts, with (on average) red-shifts being observed in the
supergranulation network regions and smaller red-shifts being observed in cell
interior regions (Peter, 2000a,b). (Blue-shifts are also observed at 
individual locations- see Section 4.) Integrated over a star, the combined
 effects of
these motions will contribute to the observed line widths. Peter
(2000a,b) has made quantitative estimates of this effect by studying    
widths and shifts from relatively small regions with those in the mean
spectrum  observed over a much larger area. Using Peter's (2000b) 
corrected widths, the values of $\xi$ deduced from the mean spectra
would overestimate the median values by about 9~per cent (narrow
component) to 15~per cent (broad components). (Two-component fits to
our spectra are discussed in Section 5). For all except the stronger
(alkali-like) lines, these potential contributions are less than our
estimated uncertainties in the line widths. 
  
Assuming that the line widths are due to non-thermal motions they can
be used to estimate the total non-thermal kinetic energy density in the
gas, $U_{\mbox{\scriptsize NT}}(T_{e})$, which is given by

\begin{equation}
U_{\mbox{\scriptsize NT}} = \frac{3}{4} \rho \xi^2
\end{equation}
where $\rho$ is the mass-density. This follows from the assumption that the 
non-thermal velocities follow a Maxwell-Boltzmann distribution in three 
dimensions, with most probably speed $\xi$. (See the Appendix for the cases 
of motions in one and two dimensions).  $\rho$ is given by

\begin{equation}
\rho = m_{\mbox{\scriptsize H}} \mu N_{g} 
\end{equation}
where $m_{H}$ is the mass of hydrogen, $N_{g}$ is the number density
of all particles and $\mu$ is the mean atomic weight defined as

\begin{equation}
\mu = \frac{\sum_{i} m_{i} N_{i}}{m{\mbox{\scriptsize H}} \sum_{i} N_{i}}
\end{equation}
where $m_{i}$ and $N_{i}$
are the mass and number density of particles of species $i$ in the plasma and
the sums run over all such species.
Adopting abundances similar to those in the solar photosphere, 
$\mu$ can be approximated by

\begin{equation}
 \mu \approx \frac{1.4 N_{\mbox{\scriptsize H}}}{N_{\mbox{\scriptsize e}} +
 1.1 N_{\mbox{\scriptsize H}}}
\end{equation}
 where $N_{\mbox{\scriptsize H}}$ is the number density of hydrogen
 and $N_{\mbox{\scriptsize e}}$ is the electron number density.
 This gives $\mu N_{g} \approx 1.4 N_{\mbox{\scriptsize
H}}$. Hence

\begin{equation}
U_{\mbox{\scriptsize NT}} \approx 1.1 m_{\mbox{\scriptsize H}} 
N_{\mbox{\scriptsize H}} \xi^2~.
\end{equation}

Fig. 2 shows a plot of $U_{\mbox{\scriptsize NT}}$ in the gas against
temperature. $N_{\mbox{\scriptsize H}}$ has been taken from a 
semi-empirical model of the average atmosphere of 
$\epsilon$~Eri (Sim 2002). The model is based on an emission measure 
distribution derived from line fluxes observed with STIS and the 
chromospheric model of Thatcher, Robinson \& Rees (1991). This new
model, and others, will be discussed in a forthcoming paper (Sim, in 
preparation). The important parameters in the model adopted here
(column mass density,
electron temperature, electron pressure, proton pressure, hydrogen 
pressure and sound speed)
are shown in Fig. 3.
The error bars in Fig. 2 give the 
total uncertainty in the values of $\xi$, but not the uncertainties in
$\rho$. The form of $U_{\mbox{\scriptsize NT}}$ with $T_e$ is
dominated by the variation of $\rho$, and is therefore considerably
more sensitive to the adopted atmospheric model than to the measured
values of $\xi$ ($\rho$ varies by more than 
five orders of magnitude across the temperature range in Fig. 2, 
compared to only a factor of three variation in $\xi$).
It is clear, however, that the non-thermal energy density in the
chromosphere is at least three orders of magnitude greater 
than in the overlying transition region and corona.

\begin{figure}
\epsfig{file=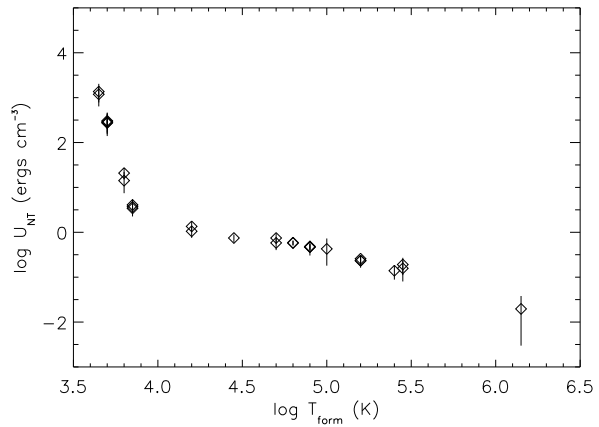, width=8.6cm}
\caption{The non-thermal energy density in the gas plotted as $\log U_{\mbox{\scriptsize NT}}$ versus $\log T_{\mbox{\scriptsize form}}$. The vertical bars show the uncertainty in $U_{\mbox{\scriptsize NT}}$ due to the total uncertainty in $\xi$, but make no allowance for possible errors in $\rho$.}
\end{figure}

\begin{figure*}
\epsfig{file=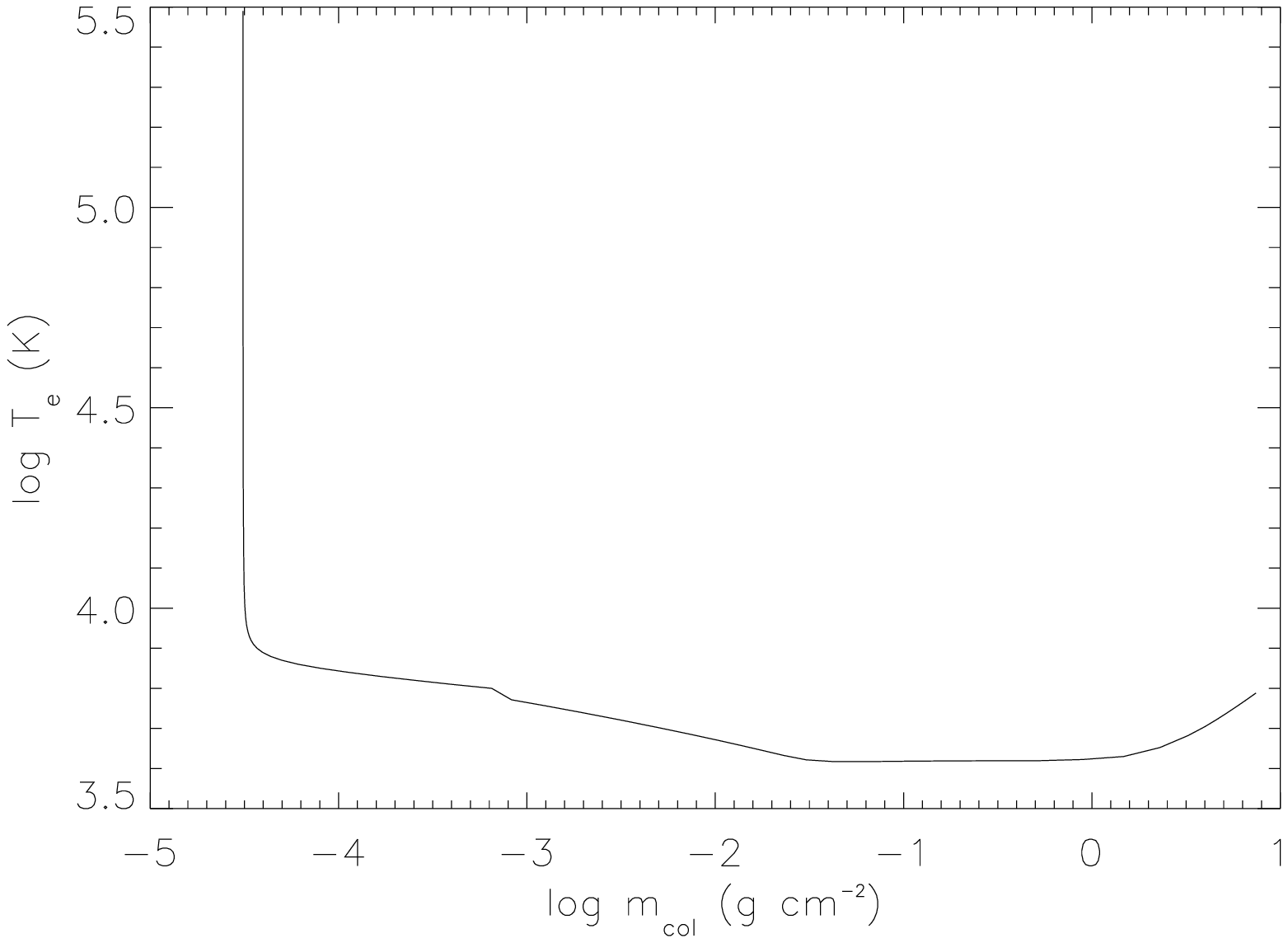, width=8.6cm}
\epsfig{file=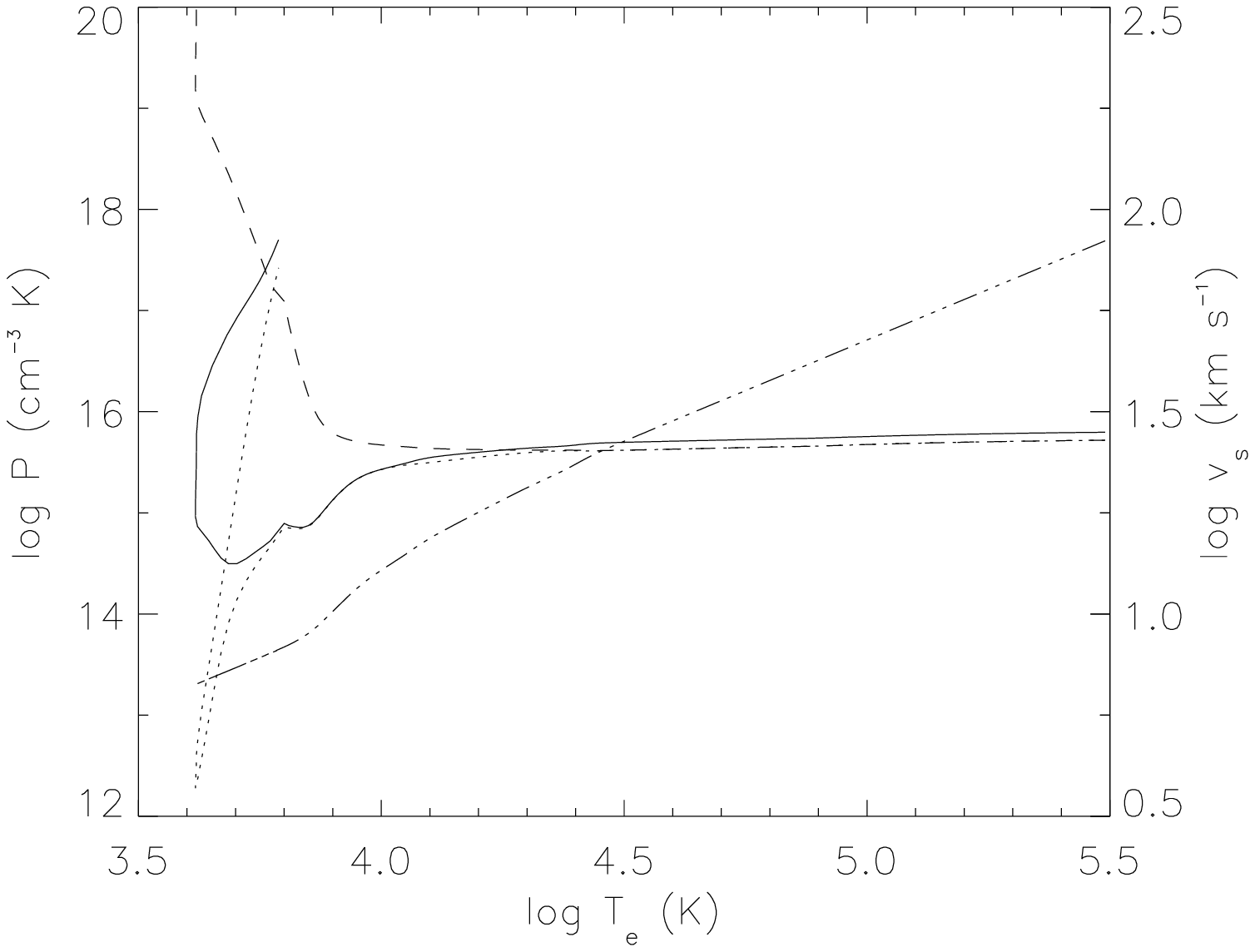, width=8.6cm}
\caption{The adopted atmospheric model. The left-hand panel shows the
electron temperature ($T_{\mbox{\scriptsize e}}$) versus the mass column 
density ($m_{\mbox{\scriptsize col}}$). The right-hand panel shows the
electron (solid line), proton (dotted line) and total hydrogen (H and H$^+$,
dashed line) pressures, and also the sound speed ($v_{\mbox{\scriptsize s}}$,
triple-dot dashed line) 
versus $T_{\mbox{\scriptsize e}}$.}
\end{figure*}

When considering the propagation of waves which are coupled to the magnetic 
field (torsional Alfv\'en waves and fast mode waves when the Alfv\'en 
speed greatly exceeds the sound speed), 
the degree of ionization of the gas needs to be considered. This is
relevant to the chromospheric regions where hydrogen is not fully
ionized (see e.g. discussion in 
Osterbrock 1961). If the frequency of collisions between ions and neutrals
is much larger than the wave frequency then there will be 
collisional coupling between the ions and neutral atoms, which can cause 
damping of the waves. In this case the gas density $\rho$ is
appropriate.  Conversely, if the collision frequency is much less than
the wave frequency, the ion density $\rho_{ion}$
becomes the relevant density.  The chromospheric model
has been used to calculate $\rho_{ion}$, allowing
for the protons, `metals' 
with low first ionization potential, and the electrons produced by these and 
hydrogen (i.e. the ions and electrons are treated as a single fluid). 
$\rho_{ion}$ is then given by
\begin{equation}
\rho_{ion} = m_{\mbox{\scriptsize e}} N_{e} + m_{\mbox{\scriptsize p}}
N_{p} + m_{m}(N_{e} - N_{p})
\end{equation}
where $m_{\mbox{\scriptsize e}}$ and $m_{\mbox{\scriptsize p}}$ are
the masses of the electron and proton, respectively,
$m_{m}$ is the mean mass of ions of elements other than hydrogen divided by 
the mean ionisation stage of such ions, and 
$N_p$ is the proton number density. $\rho_{ion}$ then replaces $\rho$ in 
equation (4). 

\begin{figure}
\epsfig{file=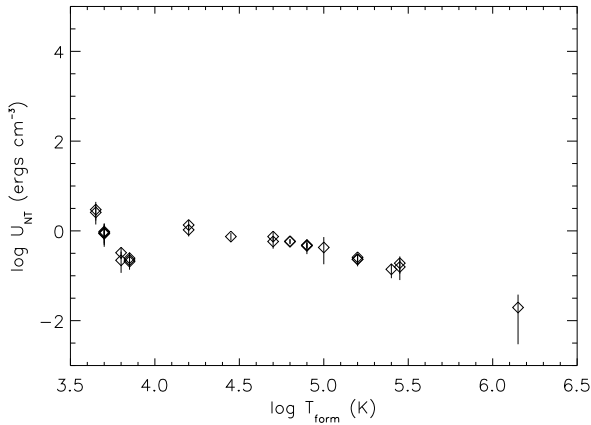, width=8.6cm}
\caption{The non-thermal energy density in the ions plotted as $\log U_{\mbox{\scriptsize NT}}$ versus $\log T_{\mbox{\scriptsize form}}$. The vertical bars show the uncertainty in $U_{\mbox{\scriptsize NT}}$ due to the total uncertainty in $\xi$, but make no allowance for possible errors in $\rho_{ion}$.}
\end{figure}

Fig. 4 shows the non-thermal energy density in the ions, which at $T_{e} \le 
10^4$~K is significantly smaller than the total non-thermal energy density 
shown in Fig. 2. Since collisions between the ions and neutrals will
occur and will transfer some energy to the neutrals, the actual
relevant non-thermal energy density will lie between the extremes  
given by Figs. 2 and 4. 

\subsection{Wave Heating}

We now consider the possibility that the non-thermal energy density is
associated with waves propagating at either the Alfv\'en speed or the sound 
speed, and examine the extent to which these hypotheses can account for the 
observations. 

Provided the wave frequencies are sufficiently small (so that the WKB 
approximation is valid), and non-linear effects are small, the non-thermal 
energy flux, $F_{\mbox{\scriptsize NT}}$, is related to the kinetic
energy density by

\begin{equation}
F_{\mbox{\scriptsize NT}} = 2 U_{\mbox{\scriptsize NT}} 
v_{\mbox{\scriptsize prop}}  
\end{equation}
where  $v_{\mbox{\scriptsize prop}}$ is the appropriate propagation 
velocity and the factor of 2 accounts for equipartition of energy
between kinetic and potential energies in the waves. 

First we consider waves which propagate at the Alfv\'en speed, without
distinguishing between torsional (shear) waves or compressional (fast-mode)
waves when $v_{a} >> v_{s}$ (see Cross 1988 for Alfv\'en wave theory 
and references to early work). The formulation is given in terms of 
$\rho_{ion}$. 

\begin{equation}
v_{a} = \frac{B}{\sqrt{4 \pi \rho_{ion}}} 
\end{equation}
where $B$ is the magnetic flux density.
$B$ is not known directly above the photosphere. R\"{u}edi et
al. (1997) have measured the spatially averaged 
surface magnetic flux density, $B_{0} = B_{s} A_{s}/A_{*}$, where
$B_{s} A_{s}$ is the surface magnetic flux. They find a value of 
$165 \pm 30$~G. In the overlying atmosphere we make the assumption 
that some fraction of the total stellar surface area is occupied by 
flux tubes, each with the same values of $B$ and individual cross-sectional
 areas. Since not all the surface magnetic
flux may continue through to higher layers, we specify the fraction which
continue at a given $T_e$ as $b = B A/B_{0}A_{*}$, 
where $A_{*}$ is the total surface area and $A$ is the actual total surface
 area occupied by the magnetic field $B$, at  given $T_e$. The observed 
spatially averaged non-thermal flux is then

\begin{eqnarray}
F_{\mbox{\scriptsize NT}} \frac{A}{A_{*}}& = & \frac{3}{2} 
\rho_{ion} \xi^{2} v_{a} \frac{A}{A_{*}} \nonumber \\
& = & \frac{3}{2 \sqrt{4 \pi}} \rho_{ion}^{1/2} \xi^{2} b B_{0} \;  
\; \mbox{erg~cm$^{-2}$~s$^{-1}$}  
\end{eqnarray}
where $F_{NT}$ refers to an individual flux tube.
The factor 3, which was introduced in the calculation of the non-thermal
energy density by the assumption of isotropic non-thermal motions in three
dimensions, has been retained here. If the motion were in fact restricted
to only 
2 or 1 dimensions, as associated with purely transverse or longitudinal
oscillations, equation~(12) would overestimate the actual flux
by factors of 1.5 and 3 respectively.

The spatially averaged non-thermal flux calculated using 
equation (12) and the observed value of $B_{0} = 165$~G
 is shown in Fig. 5, initially adopting $b = 1$. Results using $\rho$
and $\rho_{ion}$ are illustrated. The flux required to account
for the radiation losses above a given $T_e$ is also shown. The
radiation losses have been calculated using the model shown in Fig. 3,
which is based on the spatially averaged 
emission measure distribution derived by Sim (2002) and the radiative
power losses
of Cook et al. (1989), adopting the stellar photospheric abundances (which are
similar to those of the Sun). 

\begin{figure}
\epsfig{file=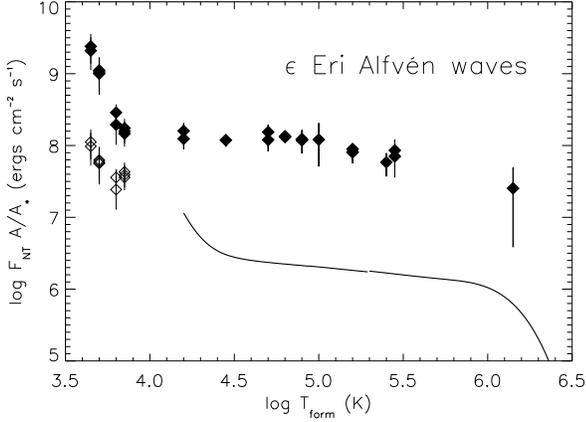, width=8.6cm}
\caption{$\log F_{\mbox{\scriptsize NT}} A/A_{*}$ versus $\log T_{\mbox{\scriptsize form}}$ for Alfv\'{e}n waves, adopting $b = 1$. The vertical error bars indicate the uncertainty in $F_{\mbox{\scriptsize NT}}$ due to the total uncertainty in $\xi$; no allowance is made for uncertainties in the values of $\rho_{ion}$ in the atmosphere. 
The open symbols indicate the calculations using $\rho_{ion}$ and the filled 
symbols show those with the total density $\rho$.
The full line shows the flux required to account for the radiation losses above a given $T_e$. }
\end{figure}

We also consider heating by acoustic waves propagating at the sound
 speed, given by

\begin{equation}
v_{s} = \left( {\frac{\gamma k_{B} T_{e}}{\mu m_{\mbox{\scriptsize H}}} } 
\right)^{1/2}  
\end{equation}
where $\gamma=5/3$ is the ratio of specific heats and $T_{e}$ is the electron
temperature. The mean molecular weight is given by equation (7).
If the acoustic waves originate in some fraction of the surface area 
($A/A_{*}$), the corresponding spatially averaged non-thermal energy flux is 

\begin{eqnarray}
F_{\mbox{\scriptsize NT}} \frac{A}{A_{*}}& = &\frac{3}{2} \rho \xi^2 v_{s} 
\frac{A}{A_{*}}\\
&\approx &
 4.1 \times 10^{-20} \xi^{2} N_{\mbox{\scriptsize H}} \sqrt{\frac
{T_{e}}{\mu}} \frac{A}{A_{*}} \; \;  {\mbox{erg~cm$^{-2}$~s$^{-1}$.}}
 \nonumber
\end{eqnarray}

Fig. 6 shows the non-thermal flux assuming that the turbulent widths are
 due to the passage of acoustic waves with $A/A_{*} = 1$, and the flux 
required to account for the radiation losses above a given $T_e$. Above 
$10^4$~K, the forms of Figs. 5 and 6 are 
very similar; this is because, apart from the numerical constant, the only
 difference between equations (12) and (14) is the factor
$\sqrt{T_{e} (N_{e} + 1.1 N_{\mbox{\scriptsize H}})} = \sqrt{P_{g}}$.
 The pressure scale height is much greater than the thickness of the 
transition region and so $P_{g}$ varies little between $\log T_{e} = 4.0$ and
 $\log T_{e}=6.0$. However $P_{g}$ increases rapidly below $T_{e} \simeq 10^4$K,
which is why Fig. 6 displays a steeper gradient in the chromosphere.

\begin{figure}
\epsfig{file=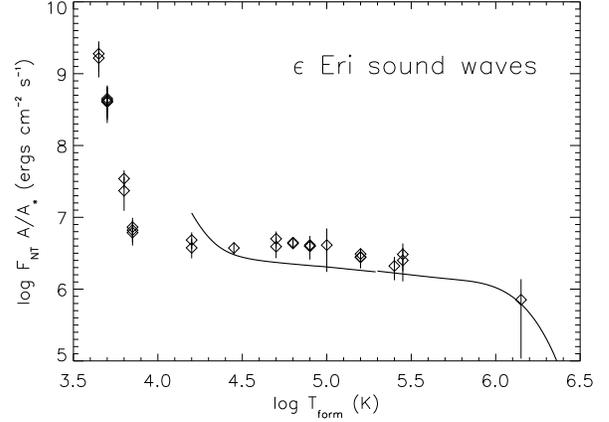, width=8.6cm}
\caption{$\log F_{\mbox{\scriptsize NT}}$ versus $\log T_{\mbox{\scriptsize form}}$ for acoustic waves. The vertical error bars indicate the uncertainty in $F_{\mbox{\scriptsize NT}}$ due to the total uncertainty in $\xi$, but do not attempt to estimate the uncertainty in the variation of $\rho$ in the atmosphere. 
The full line shows the flux required to account for the radiation losses above a given $T_e$. }
\end{figure}

The form of the curves in Figs. 5 and 6 show that in both cases the 
non-thermal wave flux is roughly constant between $\log T_{e} = 4.2$ and 4.8, 
consistent with a transport of non-thermal energy with little dissipation. 
This apparent constancy of the wave flux also shows up in Fig. 1, since over
this range of temperature $\xi$ is approximately $\propto T_{e}^{1/4}$, as in 
the solar transition region (Jordan 1991). This gradient is expected 
if there is a constant Alfv\'en  
wave flux (in the WKB approximation), or a constant acoustic flux, provided 
the electron pressure and $b$ are constant. Under the same 
circumstances a constant turbulent energy flux would give 
$\xi \propto T_{e}^{1/3}$. 
At higher temperatures, $\xi$ is roughly constant and then decreases, so 
$F_{\mbox{\scriptsize NT}}$ appears to {\it decrease}.  
This could occur if energy is being dissipated (but see below). It is not 
surprising that $\xi$ hardly varies over the mid-transition region. The 
temperature gradient in the spatially averaged models is large (as in the 
Sun), so the range of $T_e$ involved corresponds to a small height difference 
(less than 100~km). 
   
In the case of an acoustic energy flux it is possible that the decrease above 
$10^5$~K
 is simply a geometrical effect; the area occupied by the regions associated 
with the acoustic flux could {\it increase} with $T_e$.
The flux shown in Fig. 6 is the spatially averaged value, with the term 
$A/A_{*}$ on the r.h.s of equation (14) taken as 1. An increase in $A/A_{*}$
from a low value in the transition region to a larger value in the corona 
could lead to $F_{\mbox{\scriptsize NT}} A/A{_*}$ being constant until 
dissipation occurs in the corona. 

Simple energy requirement arguments can probably rule out coronal heating by 
acoustic waves.
From Fig. 6 the acoustic energy flux passing into the corona at $T_{e} \simeq
1.4 \times 10^6$~K is $\approx 8 \times 10^5$ ergs~cm$^{-2}$~s$^{-1}$, 
assuming that $A/A_{*} = 1$ by this temperature. Although this could account 
for the coronal radiation losses, it is significantly smaller than the 
estimated energy loss by thermal conduction at $1.4 \times 10^{6}$~K 
($\sim 3 \times 10^6$~erg~cm$^{-2}$~s$^{-1}$, according to Jordan et al. 
2001a). Also, there is insufficient flux below $\simeq 2 \times 10^4$~K.
Thus the simple acoustic model appears to be 
unable to provide enough energy to heat the corona of $\epsilon$ Eri.

For an Alfv\'{e}n wave flux the argument concerning the area factor fails 
because the factor $b$ in equation (12) can only be 
constant or {\it decrease} with $T_e$.  There is ample energy flux to
 account for the energy losses from the corona if $b = 1$ (i.e. all
 the surface magnetic flux extends to the corona). Indeed, only around
 20~per cent of the surface magnetic flux need extend to the corona to match
 the estimated energy flux lost by thermal conduction at, and radiation 
losses above, $1.4 \times
 10^6$~K ($\simeq 4 \times 10^6$~erg~cm$^{-2}$~s$^{-1}$). With the fluxes 
shown in Fig.~5 difficulties 
occur with the decrease in $F_{\mbox{\scriptsize NT}} A/A_{*}$ above $10^5$~K. 
This decrease could be interpreted as  being due to
 energy dissipation in the upper transition region, 
which could in principle be balanced by radiation losses and thermal 
conduction. However, including a heating process in addition to
 radiation losses will act to increase the gradient of
 the theoretical emission measure distribution (EMD) (see Jordan 2000). To
 match the decrease in $F_{\mbox{\scriptsize NT}} A/A_{*}$, apparent
 above $10^5$~K in Fig. 5, would require the gradient of the observed   
EMD to be at least an order of magnitude greater than that derived
 from line fluxes. Thus an interpretation of the fluxes in Fig. 5 in terms of 
heating in the upper transition region is not viable. If $b$ is less than 1,
then this constraint is less demanding.

 The radiation losses 
shown in Figs.~5 and 6 ignore the role played by thermal conduction above 
$T_{e} \simeq 2 \times 10^5$~K. If the thermal conduction leaving  
the corona is balanced by the local radiation losses, then no additional 
heating is required between $2 \times 10^5$~K and the base of the corona.
Over this temperature range, when comparing with the heating flux, one needs 
to consider only the 
conductive flux and radiation lost from the corona, with further radiation 
losses being added below $2 \times 10^5$~K.        
 
\begin{figure}
\epsfig{file=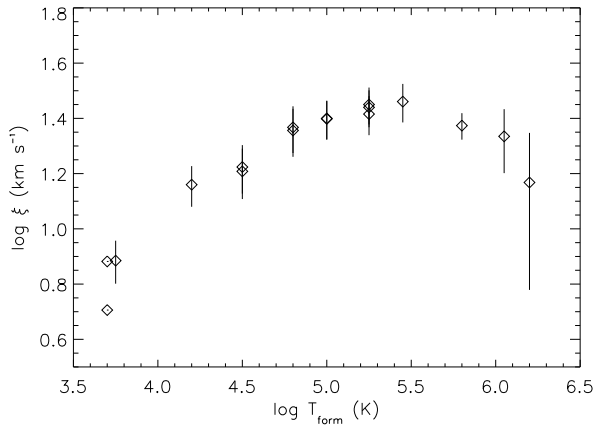, width=8.6cm}
\caption{$\log \xi$ versus $\log T_{\mbox{\scriptsize form}}$ from the solar data of Chae et al. (1998).}
\end{figure}

The heating requirements of the chromosphere can also be
considered. First we consider an acoustic flux. The 
form of Fig. 6 shows that if the turbulent velocities are associated 
with propagating waves a substantial fraction of the total non-thermal energy
 would be deposited in the low chromosphere.
The non-thermal energy flux would have to decrease by 
$\sim 10^9$ ergs~cm$^{-2}$~s$^{-1}$ across the low chromosphere. This far
exceeds any reasonable estimate of the radiative losses from this region.
 Linsky \& Ayres (1978) estimated that in cool main-sequence stars, the
 radiative fluxes in the Mg {\sc ii} h \& k lines are comparable to 30 per 
cent of the total radiative flux from the chromosphere. The total observed 
flux at the Earth in the Mg {\sc ii} h and k lines in $\epsilon$~Eri (in our
 STIS spectra) is around $5.4 \times 10^{-11}$~erg~cm$^{-2}$~s$^{-1}$, 
 corresponding to a stellar surface flux of $1.7 \times 10^6$~erg~cm$^{-2}$~
s$^{-1}$. This gives an estimated total surface radiative flux from the 
chromosphere of $8.8 \times 10^{6}$~ergs~cm$^{-2}$~s$^{-1}$. Although only a 
crude estimate, this shows that far too much energy would be dissipated in the
 low chromosphere if the observed non-thermal motions are interpreted in terms
of a propagating acoustic wave flux.  

If we invoke an area filling factor, then in the chromosphere, $A/A_*$ would
have to be much lower than at $10^4$~K, and would need to increase rapidly by
around $10^4$~K. This seems unlikely, as in the Sun, acoustic waves are 
thought to be important only in the interiors of supergranulation cells, whose
area does not change significantly in the overlying regions. 
By analogy with the solar chromosphere, the spatially integrated emission and 
the observed line widths are
likely to be dominated by the contribution from the supergranulation 
boundaries. In this case, relating the line widths to acoustic waves will not
be correct, and the apparent discrepancy between the observed and deduced
fluxes is understandable.  
   
From the limits to the Alfv\'en wave flux shown in Fig. 5 
it is clear that there is no difficulty in principle in obtaining
sufficient flux to
 pass through to the lower transition region and the flux around 
$2 \times 10^{4}$~K represents an upper limit available to the overlying 
regions. We do not address the question 
of how the supergranulation boundary regions in the chromosphere are heated.
(Osterbrock 1961 gives a discussion of the possible processes involved, but 
with numerical values of parameters that require revision in the light of 
modern measurements of magnetic fields and current chromospheric models).  

In Fig. 7 we show the non-thermal velocities calculated using the
solar line widths given by Chae et al. (1998). Fig. 8 shows the
non-thermal wave fluxes calculated from
 equations (12) and
 (14) using these non-thermal velocities and densities 
taken from model C (average quiet Sun) of Vernazza, Avrett, \& Loeser (VAL)
 (1981). Again, the upper and lower limits on the Alfv\'en wave fluxes
are shown. As in Fig. 5, $b$ has been set to 1.0. The adopted temperatures of 
line formation are similar to those used by Chae et al. (1998), except for
 the neutral chromospheric lines. For these lines 
significantly lower values of $T_{\mbox{\scriptsize form}}$ have been used,
 based on radiative transfer calculations with the VAL-C (1981) model.
 These lower temperatures lead to a difference in the dependence of the fluxes
with $T_e$, when comparisons are made with the calculations by Chae et al. 
(1998) (their fig. 13). The mean magnetic flux density was taken as 20~Gauss 
(see discussion in Montesinos \& Jordan 1993). Apart from the
 absolute scales, the variations of the spatially averaged fluxes are quite 
similar to those found for $\epsilon$ Eri: for Alfv\'en waves the lower limit
 to the chromospheric flux is similar to the flux in the lower transition 
region, while for acoustic waves and the upper limit to the Alfv\'en
wave flux, there is
 a large drop in $F_{\mbox{\scriptsize NT}}$ between the chromosphere and
 transition region. In both cases the flux is almost constant through the 
low transition region with a decrease in the upper transition region.
 In the Sun, the factor by which the flux decreases in the upper transition 
region is larger than in $\epsilon$~Eri, 
but begins at a slightly higher temperature. The presence of a dip at around 
$\log T_{e} = 4.5$  relies heavily on the accuracy of the non-thermal 
contribution to the line-width of the optically thick C~{\sc ii} lines 
deduced by Chae et al. (1998).

\begin{figure*}
\epsfig{file=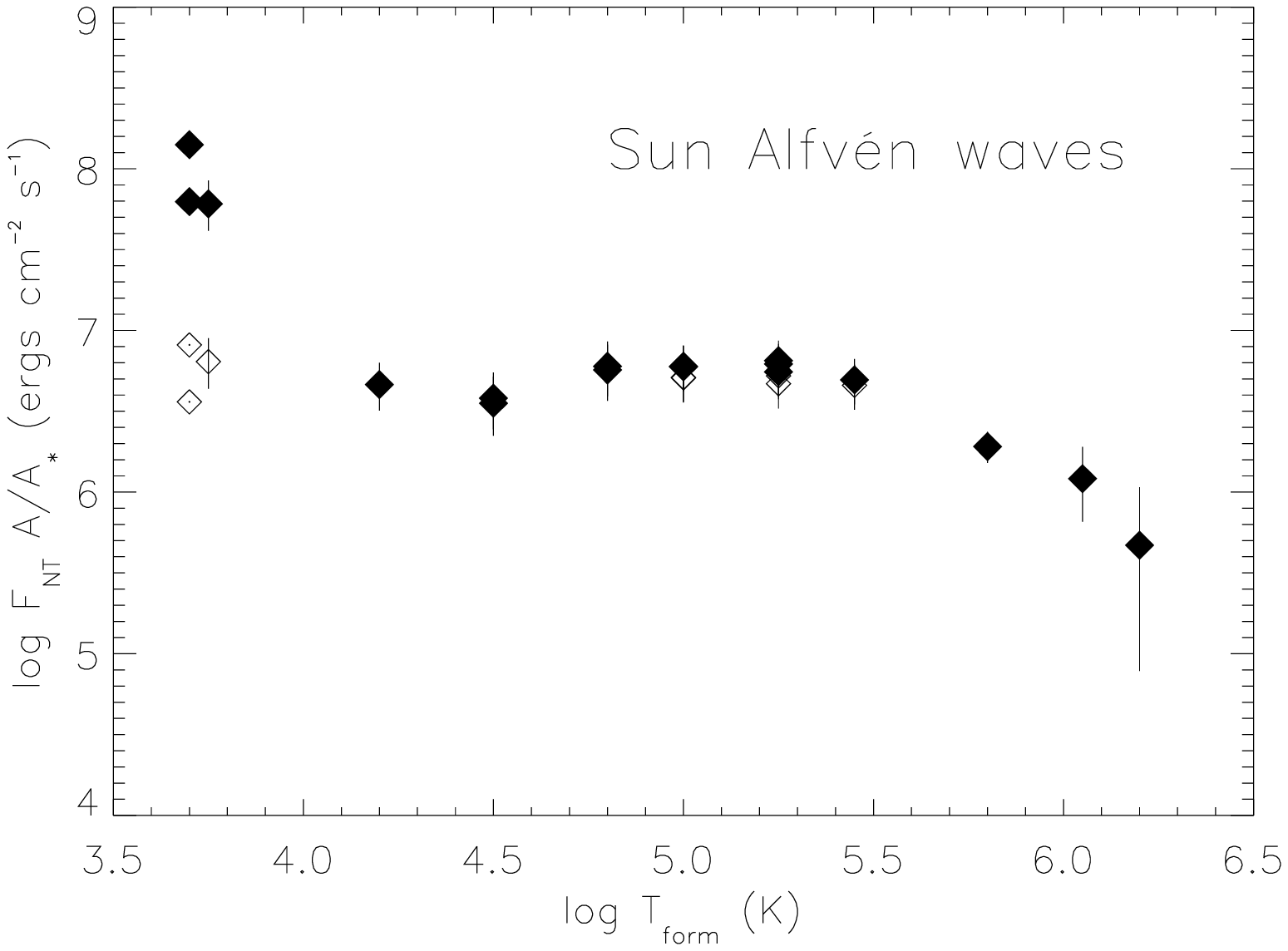, width=8.6cm}
\epsfig{file=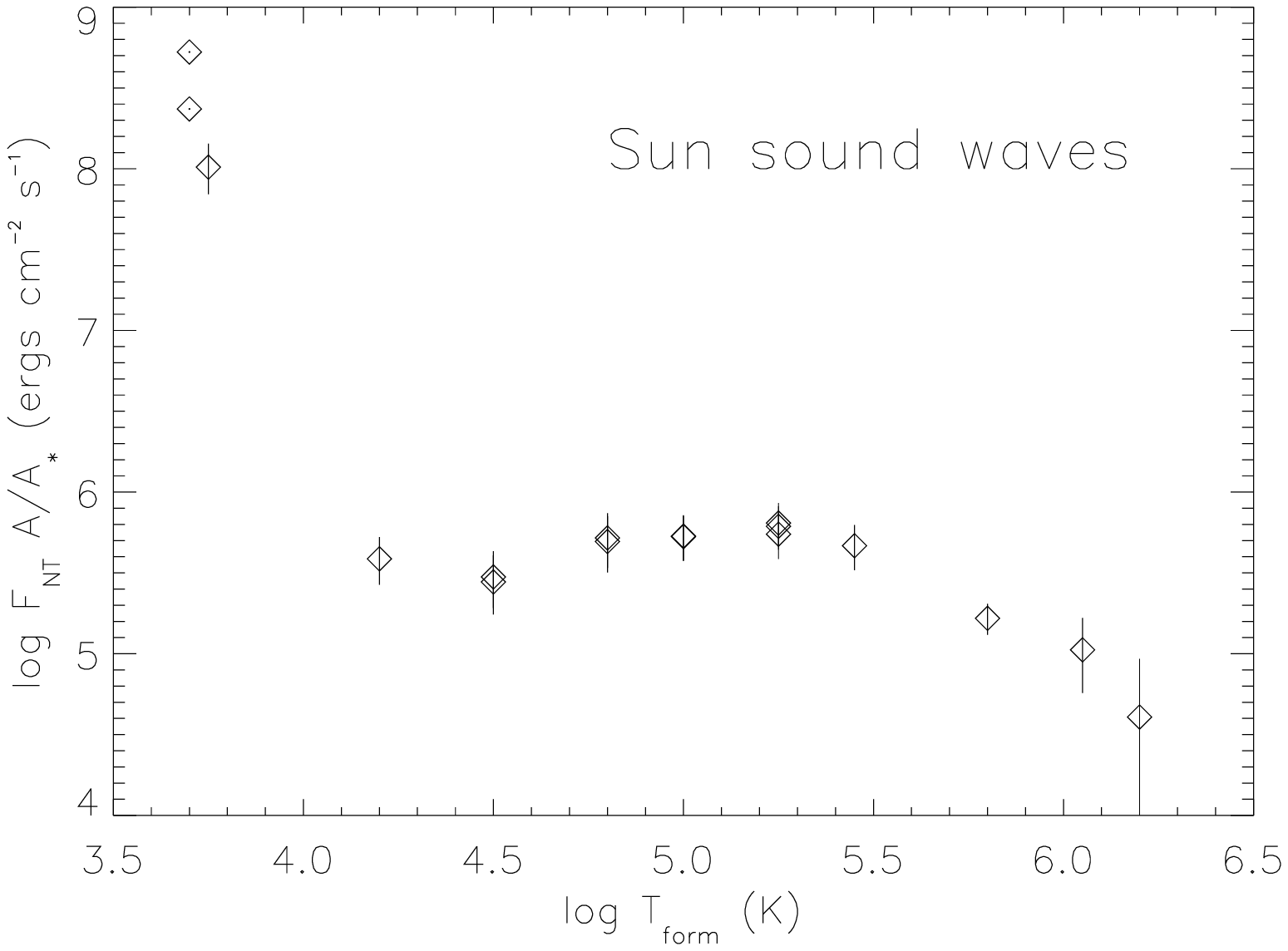, width=8.6cm}
\caption{$\log F_{\mbox{\scriptsize NT}}$ versus $\log T_{\mbox{\scriptsize form}}$ from the solar data of Chae et. al (1998). The left-hand panel is for Alfv\'{e}n waves and the right-hand panel is for acoustic waves. In the left-hand
panel the open symbols indicate the calculations using $\rho_{ion}$ and the 
filled symbols show those with the total density $\rho$.
}
\end{figure*}

It is surprising that in the Sun, and in $\epsilon$~Eri, in the low transition
 region $\xi$ increases according to $T_{e}^{1/4}$, consistent with a constant
Alfv\'en
wave energy flux in the WKB approximation. This approximation is valid only 
if the wavelength is less
than the scale-height of the Alfv\'en wave speed (see e.g. Meyer 1976; 
Hollweg 1991 and references therein); i.e. if the Alfv\'en wave period 
$\tau_{w}$ is such that
\begin{equation}
\tau_{w} << \frac{2 \pi }{v_a} 2 H  
\end{equation}
where $H$ is the scale-height of the Alfv\'en speed. If $\tau_{w}$ is much
larger than this value, then the waves become evanescent and $\xi$ would be
constant with height (which is observed to occur around $10^5$~K).   
Defining $H = v_{a} (\mbox{d}h/\mbox{d}v_{a})$, the critical period can be expressed as
\begin{equation}
\tau_{c} = \frac{2.0 \times 10^{-10} Em(0.3)_{t} T_{e}^{3/2}}{B P_{e}^{3/2}
(0.5 + \frac{\mbox{\scriptsize d} \log B}{\mbox{\scriptsize d} \log T_{e}})} 
\end{equation}
where the temperature gradient has been written in terms of the true 
(intrinsic) emission measure and the electron pressure 
($P_{e} = N_{e} T_{e}$), i.e.
\begin{equation}
 Em(0.3)_{t} = \frac{P_{e} P_{H}}{\sqrt {2} T_{e}} \frac{\mbox{d}h}{\mbox{d}T_{e}}~.  
\end{equation}
For a plane parallel layer $Em(0.3)_{t}$ can be replaced by
$2 Em(0.3)_{obs} A_{*}/A$. As before, $B A$ is replaced by $B_{o} A_{*} b$,
 and the unknown area A, but not its variation with $T_e$, cancels.  
Working above $\simeq 2 \times 10^4$~K (since we take $P_e$ and $\mu$ to be
 constant) the critical period can 
be found as a function of $T_e$, provided assumptions are made about $b$ and
 the change in the area factor $A$. I.e.
\begin{equation}
\tau_{c} =  \frac{2.0 \times 10^{-10} Em(0.3)_{obs} T_{e}^{3/2}}
{B_{o} b P_{e}^{3/2} [0.5 + \frac{\mbox{\scriptsize d}\log (b/A)}{\mbox{\scriptsize d} \log T_{e}}]}~.  
\end{equation}
Since the area occupied by the solar
supergranulation network varies little through the low to mid transition 
region, we assume that $A$ is constant and we have to assume that
$b = 1$. For the Sun, using $B_{0} = 20$~G,  
$P_{e} = 2.1 \times 10^{14}$~cm$^{-3}$~K, and the network
values of $Em(0.3)_{obs}$ from Macpherson \& Jordan (1999), the values of 
$\tau_c$
are $\simeq 20$~s at $2 \times 10^4$~K, $\simeq 4$~s at $8 \times 10^4$~K, and 
increase again to $\simeq 30$~s at $3 \times 10^5$~K. For the assumed 
parameters these are minimum values
since only a small increase in $A$ with $T_e$ will increase $\tau_c$, and if 
$b$ is less than 1, and decreases with $T_e$, this will also increase 
$\tau_c$. Also, the value of $B_0$ used could be a factor of 2 smaller.
Thus although the critical wave periods are short, it is not certain that
WKB waves can be excluded. The most stringent condition occurs around 
$8 \times 10^4$~K, where (maybe coincidentally) $\xi$ begins to increase by a
 power of $T_e$ which is less than 1/4. In $\epsilon$~Eri, the values of 
$Em(0.3)_{obs}$, $P_e$
 and $B_0$ are all larger, but result in values of $\tau_c$ that are smaller
by a factor of about $130$, with the same trends with $T_e$. The values of
$b$ and $\mbox{d} \log (b/A) /\mbox{d} \log T_e$, which act to increase the 
critical period,  therefore become the crucial factors in 
deciding whether or not the waves can satisfy the WKB approximation.

As Hollweg (1991, and references therein) has stressed, if the WKB 
approximation does not hold, then the energy flux carried can be larger or 
smaller than that deduced by using the approximation. There is also the 
possibility of wave reflection by the steep Alfv\'en
 wave gradient, and wave propagation at particular frequencies (resonances).
 Together, in principle, these effects might account for the
apparent decrease in the Alfv\'en wave flux that appears in Figs. 5
and 8, without invoking wave dissipation in the upper transition region.
If wave reflection occurs, there is also the possibility that eventual wave
dissipation could contribute to the heating of the lower transition region.
Although we do not attempt an analysis of possible wave processes,  
the new observations of $\epsilon$~Eri, and the new models being produced, 
will place strong constraints on future
detailed calculations of wave propagation through the transition region.  

\section{Line shifts}

Unless otherwise stated, the rest wavelengths given in Table 3 are
taken from Kurucz \& Bell (1995). In general, these agree with those
given by Kelly (1987) to within $0.001$~\AA. In Jordan et al. (2001a)
we noted that the wavelength interval between the two lines of 
N\,{\sc v} is slightly larger than that given by the values from Kelly
(1987) (which are the same as in Kurucz \& Bell 1995, and originate from 
laboratory measurements by Hallin 1966). There has been  
previous debate concerning the N\,{\sc v} wavelengths in the Sun (see 
Achour et al. 1995; Brekke, Hassler \& Wilhelm 1997, and references therein). 
Brekke et al. (1997) find that for the 1242.8-\AA~line the best
agreement is with the 
measurement by Hallin (1966) (1242.804~\AA). For the 1238.8-\AA~line, 
they find best agreement with the earlier measurement by Edl\'en (1934) 
(1238.800~\AA). Combining these wavelengths gives a wavelength interval
of 4.004~\AA. Of the various combinations of the laboratory
wavelengths this one does give the best agreement with our interval of
3.994~\AA, which should be accurate to within 0.007~\AA, and appears
to be the best measurement of the interval to date. Since we do not
have absolute wavelengths, we use the values by Hallin (1966) in Table
3, but point out that the 1242.804-\AA~line is likely to give more accurate
wavelength shifts. Our measured interval suggests that the wavelength of
the other line should be 1238.810~\AA.
 It would be
useful to remeasure these wavelengths in the laboratory with
modern techniques.  

New interferometric measurements of lines from the alkali-like ions
C\,{\sc iv} and Si\,{\sc iv} are available from the work of Griesmann
and Kling (2000), and these are adopted in Table 3. For C\,{\sc iv}
the new values agree more closely with the values of Kelly (1987) than
with those in Kurucz \& Bell (1995) and lead to smaller observed
red-shifts. The wavelength interval between the two C\,{\sc iv} lines
that we measure is definitely smaller than that given in Kurucz \&
Bell (1995), and agrees most closely with that from the wavelengths
given by Kelly (1987). For the Si\,{\sc iv} lines our measured
interval is consistent with those listed in all the above three
papers. The new measurement by Griesmann \& Kling (2000) for the line of
Al\,{\sc ii} is also adopted in Table 3, although it is not
significantly different from the earlier value, given the uncertainty
in the observed value. 

The wavelength calibration of {\it FUSE} 
is not accurate, so no absolute wavelengths can be measured for the 
O\,{\sc vi} and C\,{\sc ii}~1037-\AA~mult. lines. Since these lines
are all close together in wavelength, their relative wavelengths can
be measured. The wavelength interval between the O\,{\sc vi} lines
agrees with the values in both Kurucz \& Bell (1995) and Kelly (1987) to
within our measurement uncertainty of $0.025$\AA. That between 
 the C\,{\sc ii}   
lines agrees with the laboratory measurements for this spectroscopic
standard, to within our measurement uncertainty. There is then no
relative shift between the
O\,{\sc vi} lines and the C\,{\sc ii} lines (which are formed where the mean
 red-shift is 2--7~km~s$^{-1}$), to within the
measurement uncertainty of $\simeq 4$~km~s$^{-1}$, 
it appears that the red-shifts of the O\,{\sc vi} lines
are no greater that that of O\,{\sc v}.

The line shifts given in Table 3 are plotted as a function of wavelength 
in Fig. 9 and of temperature of formation in Fig. 10. The diamonds
indicate lines observed with the E140M grating and the asterisks those
which are observed with the lower resolution E230M grating. For multiplets 
in which the individual line shifts agree to within the measurement errors
 (see Table 3)
the individual shifts have been combined to give a single shift for the
 multiplet. 

\begin{figure}
\epsfig{file=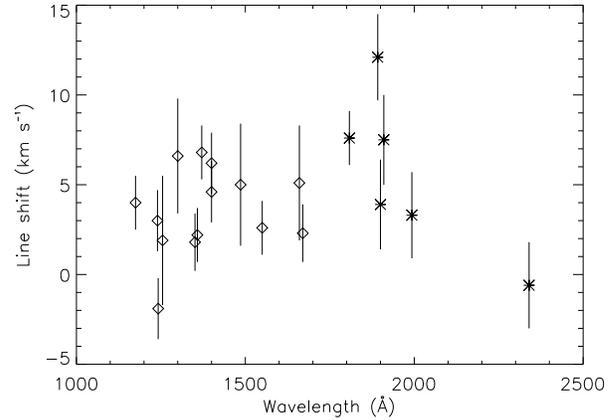, width=8.6cm}
\caption{Line shift versus wavelength. The lines observed with the E140M grating are shown by $\diamond$, and those observed with the~E230M grating by $\ast$.}
\end{figure}

Fig. 9 shows that most of the lines display a small (few km~s$^{-1}$)
 red-shift, although the scatter is quite large. There appears to be
 an offset 
in the wavelength calibration of the E230M grating (which covers the 
wavelength range $\lambda > 1750$~\AA) compared with that of
 the E140M grating, since at a given $T_e$, the lines observed around 
1800~\AA~to 1900~\AA~observed with
 the E230M grating show larger redshifts than those observed with the E140M 
grating. The offset is small and corresponds to only about 2 pixels
 in wavelength, but in the discussion below we restrict our comments to the
trends in the E140 data.

\begin{table*}
\caption{Fit parameters from two-Gaussian fits to selected strong lines. $F$ denotes the integrated line flux, given in $10^{-13}$~ergs~cm$^{-2}$~s$^{-1}$ at the Earth. The subscripts N and B denote ``Narrow'' and ``Broad'' components respectively. The errors are formal 1$\sigma$ errors deduced from the correlation matrix of each fit.}
\begin{tabular}{llllllll} \hline
 Ion& $\lambda_{0}$(\AA)&$F_{\mbox{\scriptsize N}}$ & $\Delta \lambda_
{\mbox{\scriptsize N}}$(\AA) & $\lambda_{\mbox{\scriptsize
obs,N}}$(\AA) & 
$F_{\mbox{\scriptsize B}}$ & $\Delta \lambda_{\mbox{\scriptsize
B}}$(\AA) & $\lambda_{\mbox{\scriptsize obs,B}}$(\AA) \\ \hline
C {\sc iv} & 1548.204 & $3.293 \pm 0.152$ & $0.165 \pm 0.004$ &
$1548.221 \pm 0.001$ & $2.177 \pm 0.137$ & $0.420 \pm 0.019$ &
$1548.247 \pm 0.004$ \\

C {\sc iv} & 1550.781 & $1.396 \pm 0.108$ & $0.147 \pm 0.007$ &
$1550.784 \pm 0.002$& $1.453 \pm 0.103$ & $0.372 \pm 0.016$ & 
$1550.805 \pm 0.004$ \\

N {\sc v}  & 1238.821 & $0.676 \pm 0.062$ & $0.151 \pm 0.008$& 
$1238.828 \pm 0.002$ & $0.304 \pm 0.058$ & $0.377\pm 0.047$ & 
$1238.845 \pm 0.009$ \\

N {\sc v}  & 1242.804 & $0.334 \pm 0.051$ & $0.147 \pm 0.011$ & 
$1242.817 \pm 0.003$ & $0.145 \pm 0.047$ & $0.332 \pm 0.062$ & 
$1242.865 \pm 0.028$ \\

O {\sc vi} & 1031.912 & $3.540 \pm 0.136$ & $0.173 \pm 0.003$ &
$1031.940 \pm 0.001$ & $1.065 \pm 0.127$ & $0.376 \pm 0.022$ &
$1031.959 \pm 0.005$ \\

O {\sc vi} & 1037.613 & $1.858 \pm 0.074$ & $0.170 \pm 0.004$ &
$1037.615 \pm 0.001$ & $0.417 \pm 0.067$ & $0.392 \pm 0.035$ &
$1037.630 \pm 0.006$ \\

Si {\sc iii} & 1892.030 & $0.794 \pm 0.089$ & $0.158 \pm 0.012$&  
$1892.105 \pm 0.004$ & $0.523 \pm 0.093$ & $0.427 \pm 0.042$ & 
$1892.112 \pm 0.022$ \\

Si {\sc iv} & 1393.760 & $1.285 \pm 0.050$ & $0.136 \pm 0.004$& 
$1393.780 \pm 0.001$ & $0.960 \pm 0.047$ & $0.379 \pm 0.012$ & 
$1393.793 \pm 0.003$ \\

Si {\sc iv} & 1402.773 & $0.739 \pm 0.038$ & $0.141 \pm 0.005$ & 
$1402.794 \pm 0.002$ & $0.450 \pm 0.034$ & $0.427 \pm 0.026$ & 
$1402.803 \pm 0.006$ \\ \hline
\end{tabular}
\end{table*}

\begin{figure}
\epsfig{file=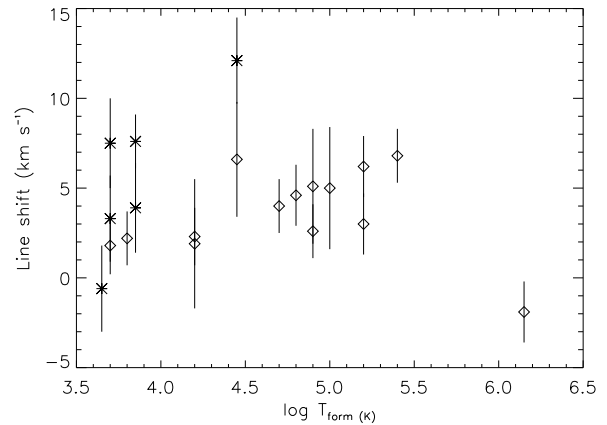, width=8.6cm}
\caption{Line shift versus $\log T_{\mbox{\scriptsize form}}$. Symbols~as in Fig. 9.}
\end{figure}

Fig. 10 does not show a strong trend in the line shifts with 
$T_{\mbox{\scriptsize form}}$. Using only the lines observed with the 
E140M grating, the mean red-shift is $3.7 \pm 0.6$~km~s$^{-1}$.
The mean red-shift of the mid transition region lines 
($4.4 < \log T_{\mbox{\scriptsize form}} < 5.5$)
is $4.9 \pm 0.8$~km~s$^{-1}$ 
while that for the lower temperature lines is
$2.1 \pm 1.1$~km~s$^{-1}$.
Thus, on average, the lower temperature lines are less red-shifted by
about $3$~km~s$^{-1}$. 
The Fe~{\sc xii} line is apparently blueshifted, but this relies on the 
solar wavelength at the limb.  

In the solar transition region, using 
spatially integrated observations at Sun-centre, Peter \& Judge (1999)
 found red-shifts of up to $9.9$~km~s$^{-1}$ in the N\,{\sc v} lines, and
 smaller red-shifts in lines formed at lower and higher temperatures. These
red-shifts decrease towards the limb. For the Si\,{\sc iv}, C\,{\sc iv} and 
N\,{\sc v} lines in common with our observations of $\epsilon$~Eri, we find 
similar red-shifts (if we adopt the wavelengths from Kurucz \& Bell 1995),
 except for the N\,{\sc v} lines, for which our values are significantly 
smaller. The new wavelengths we adopt for Si\,{\sc iv} and 
C\,{\sc iv} do, however, lead to smaller red-shifts (see Table 3). If the 
Fe\,{\sc xii} wavelength is reliable, then like Peter \& Judge (1999) we
observe a blue-shift in the Fe\,{\sc xii} line, although this is smaller in 
$\epsilon$~Eri.      

The physical origin of the 
shifts is not yet understood, although it has been suggested by Hansteen 
(1993)
that these could be due to waves propagating {\it downwards} from 
reconnection events (nanoflares) in the corona, or in the mid transition
 region (Peter \& Judge 1999).  However, our line-width
measurements, using one-Gaussian fits, show no evidence that the energy 
flux {\it increases} with 
temperature, as one would expect if the origin of the waves was within 
the corona.
In the solar photosphere the velocity fields associated with the
supergranulation flow are very small, but the possibility of convective
overshoot into the transition region, and resulting downflows in the
cell boundaries, does not appear to have been
investigated. With lower densities and mass conservation, larger
velocities are in principle possible at the level of the transition
region. It is also interesting that, in at least one solar boundary region,  
Macpherson, Jordan \& Smith
(1999) found that the downflows appear to be adjacent to,
rather than in, the brightest regions of the boundaries. 
Overall, the origin of the red-shifts is still not understood.
  
As discussed above, in stellar observations only the combined effects
of red- and
blue-shifts can be observed, which could contribute to the line
widths. The shifts observed are significant compared with the
uncertainty of $0.9$~km~s$^{-1}$ in the stellar radial velocity. The
shifts derived from two-component fits to line profiles are discussed
below in Section 5.  

\section{Two-component fits}

\begin{figure*}
\epsfig{file=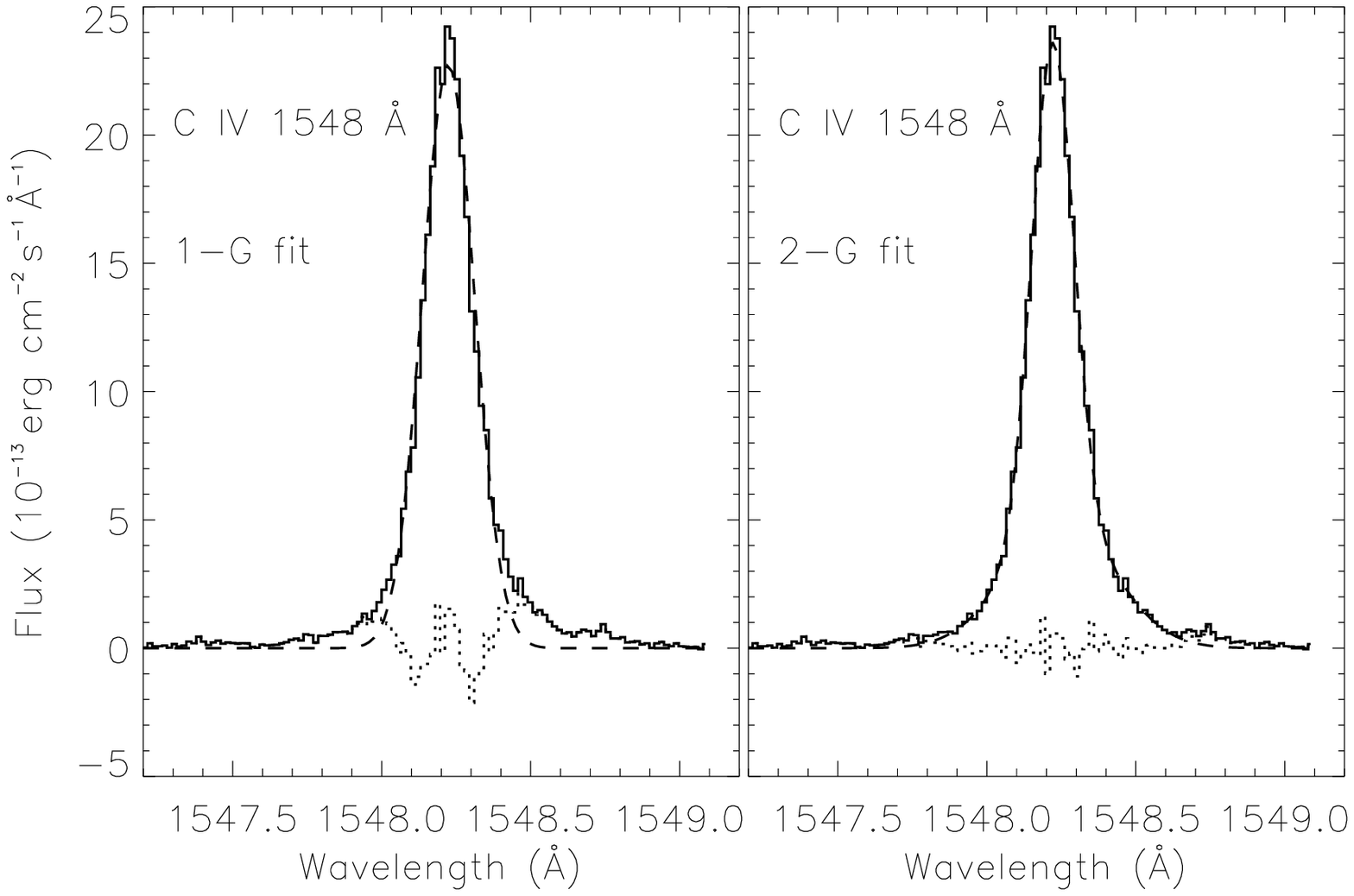,  width=8.2cm, height=4.5cm} \hspace{.5cm}
\epsfig{file=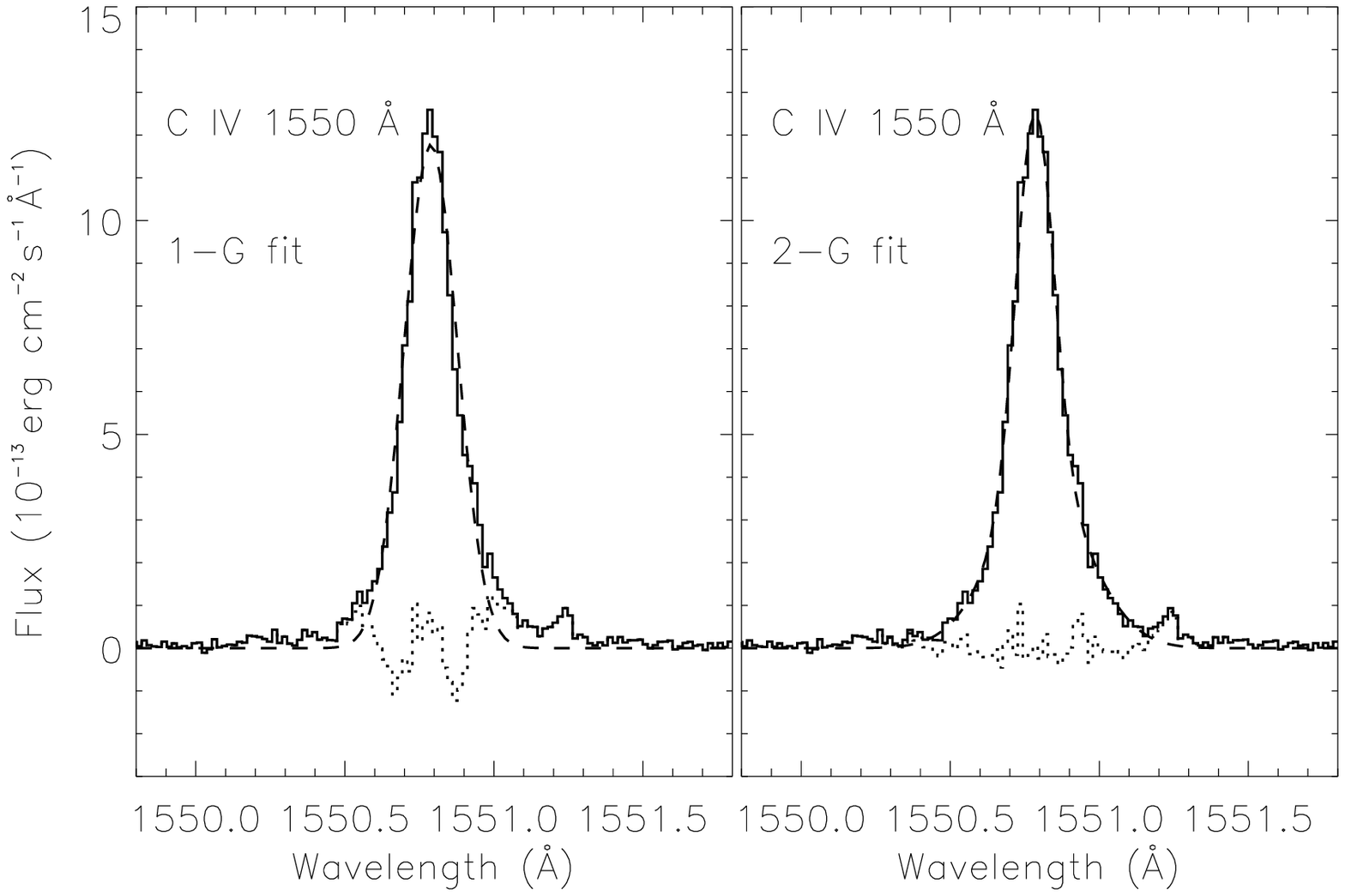, width=8.2cm, height=4.5cm}
\epsfig{file=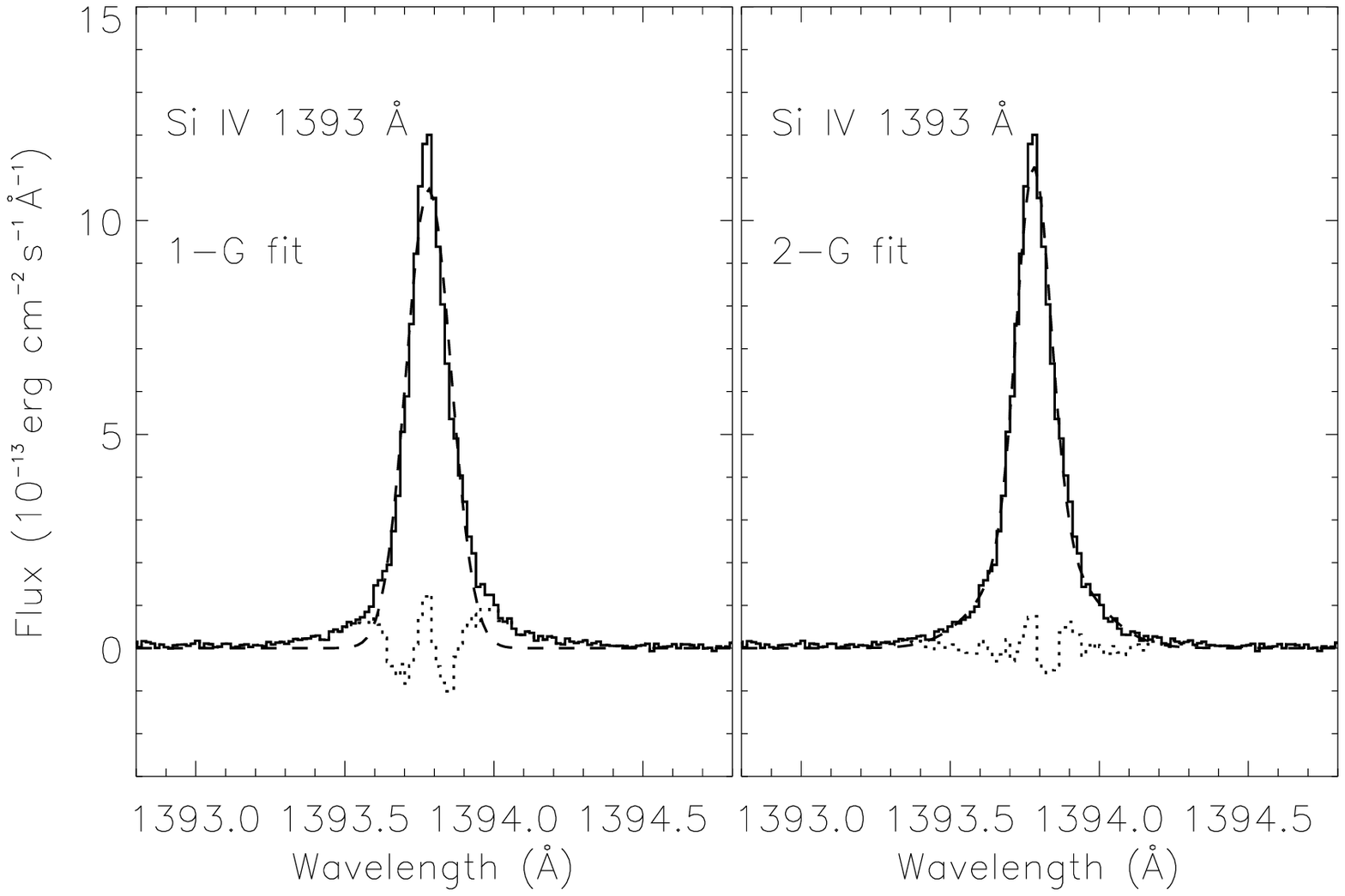, width=8.2cm, height=4.5cm} \hspace{.5cm}
\epsfig{file=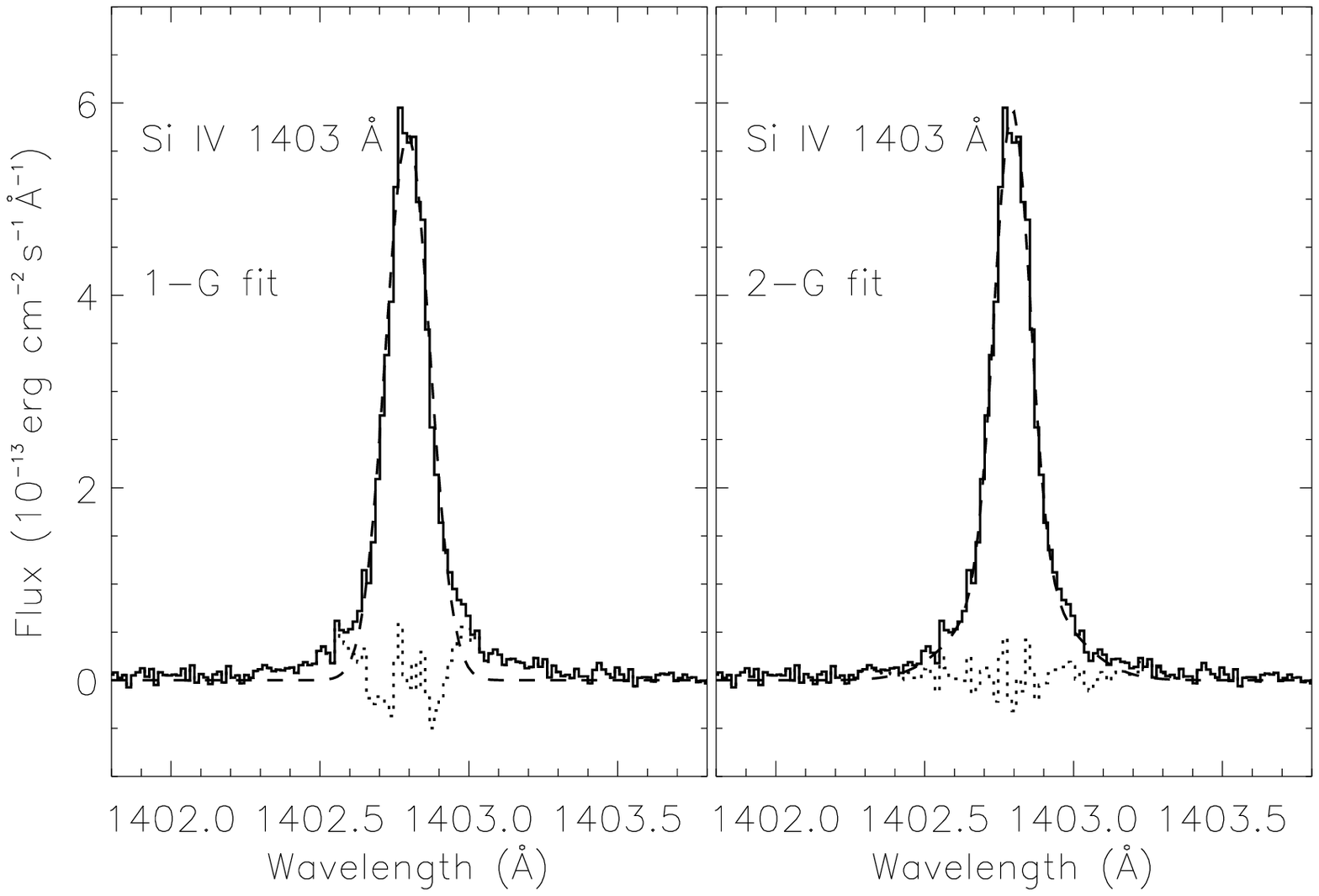, width=8.2cm, height=4.5cm}
\epsfig{file=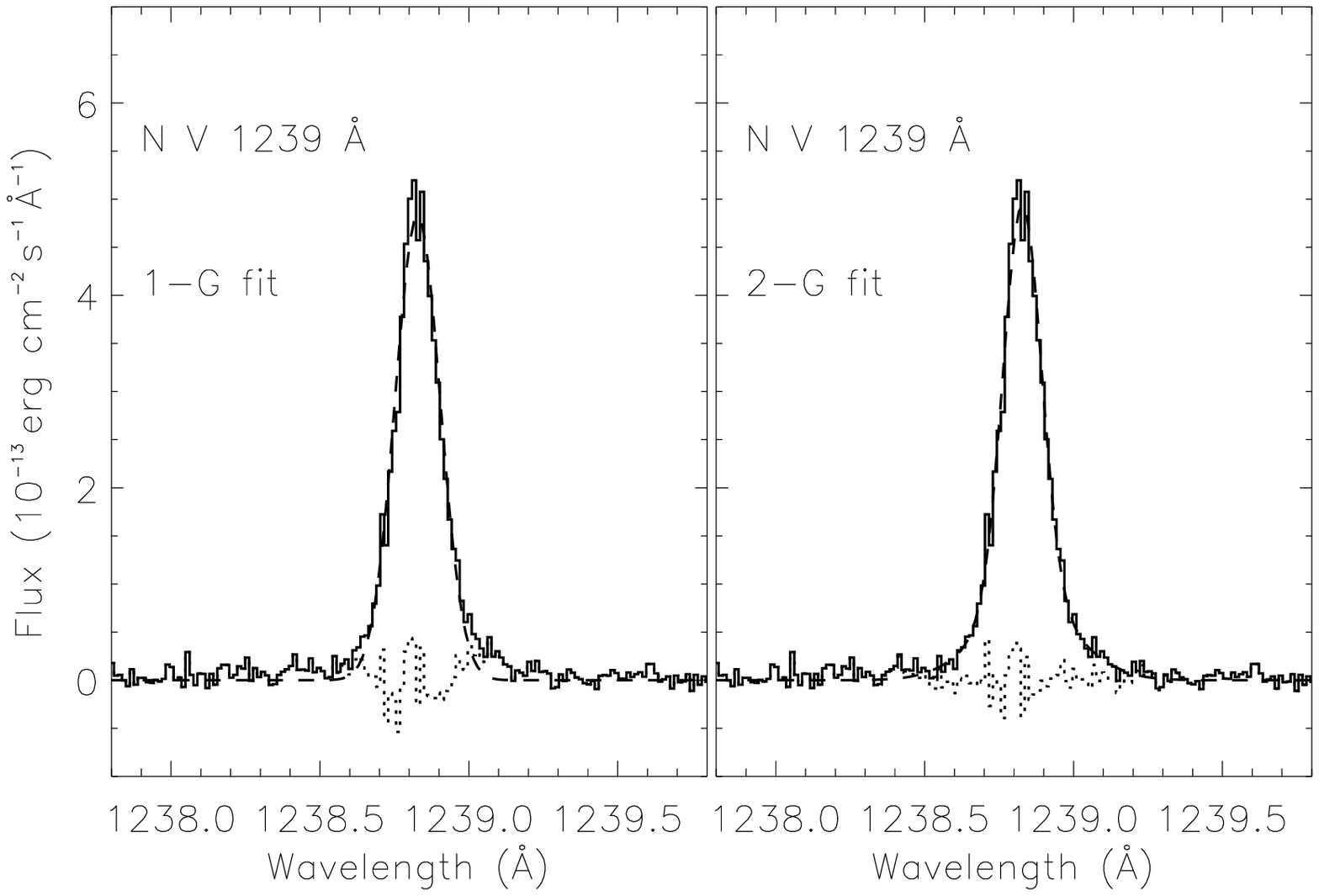, width=8.2cm, height=4.5cm} \hspace{.5cm}
\epsfig{file=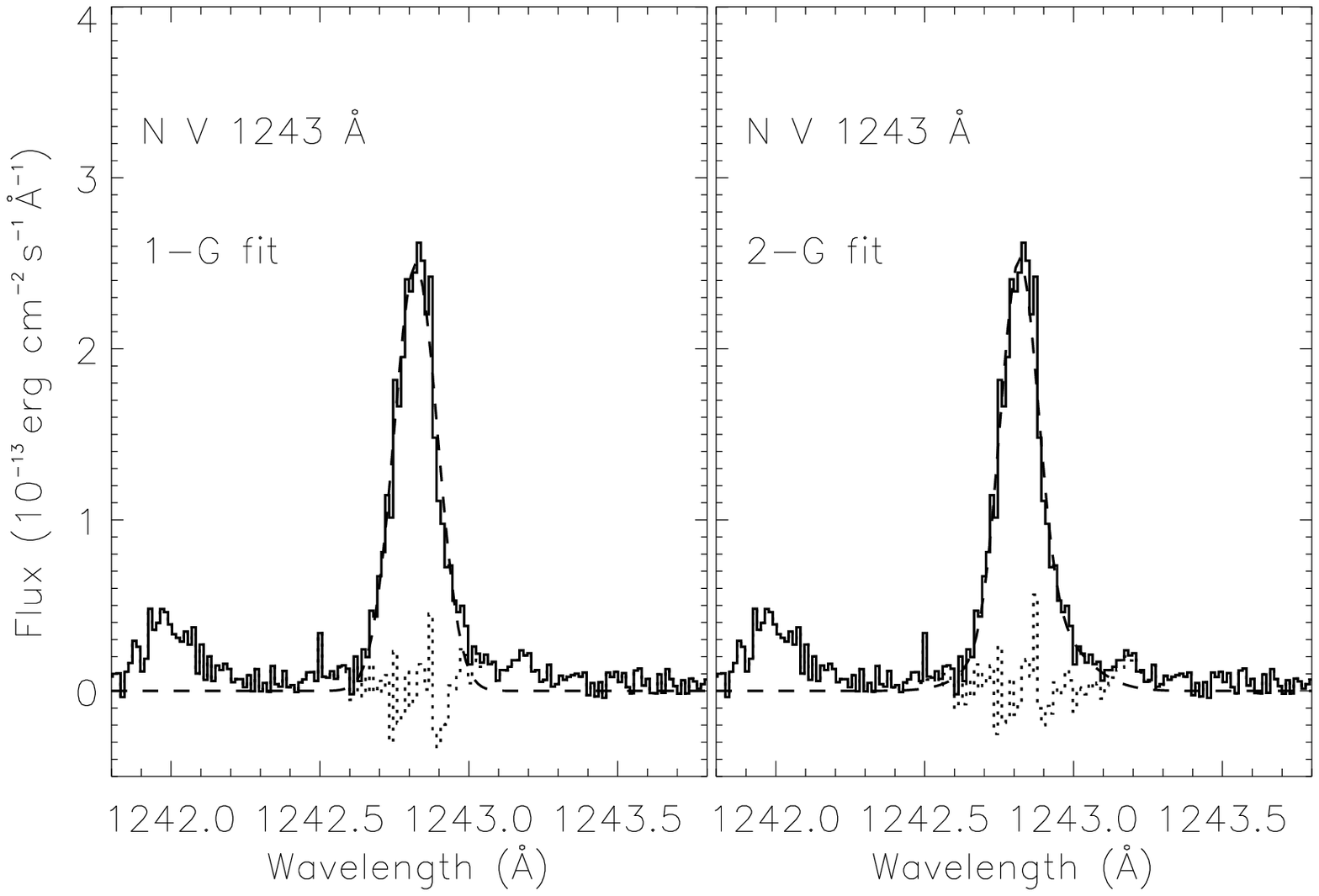, width=8.2cm, height=4.5cm}
\epsfig{file=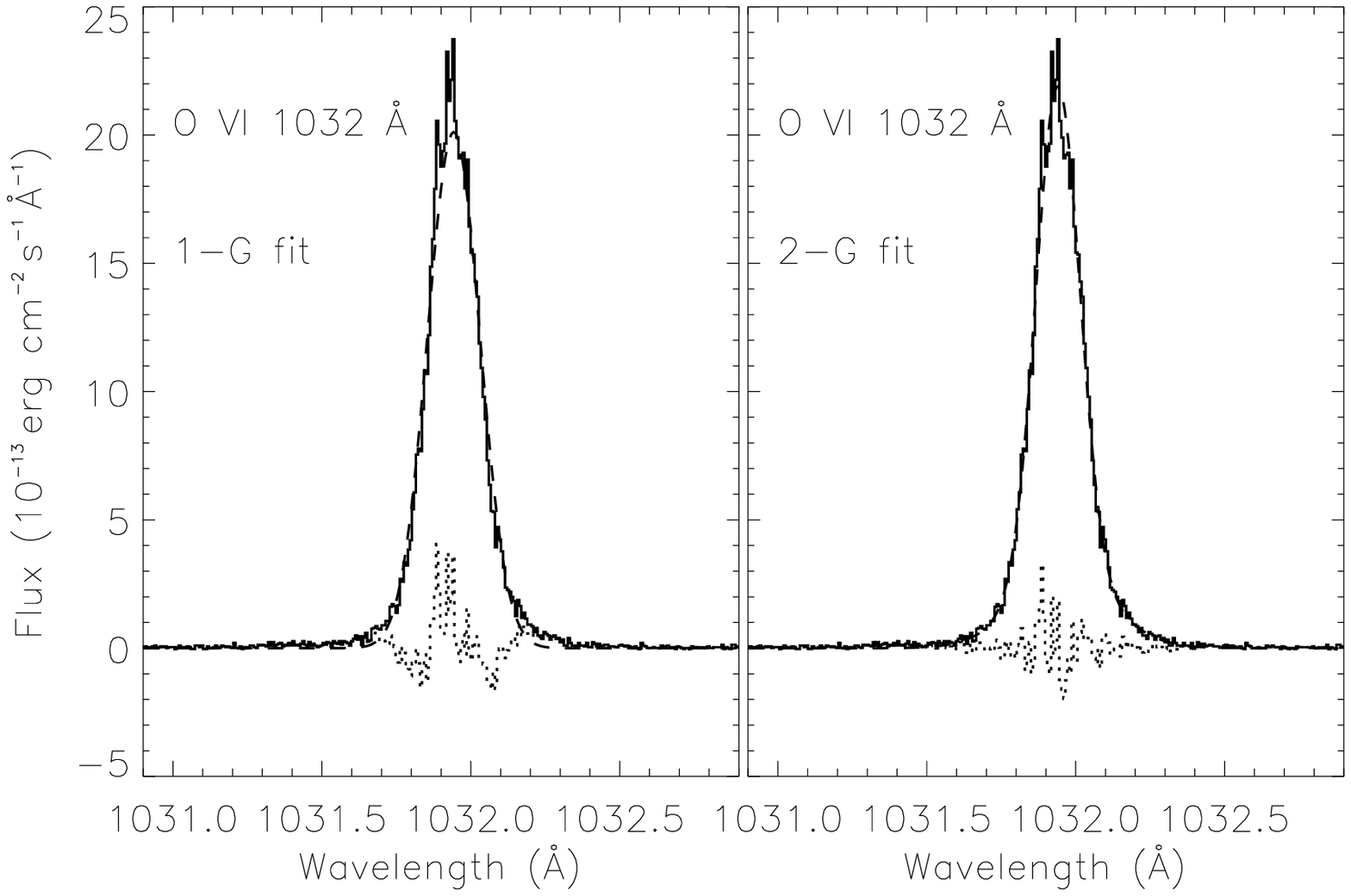, width=8.2cm, height=4.5cm} \hspace{.5cm}
\epsfig{file=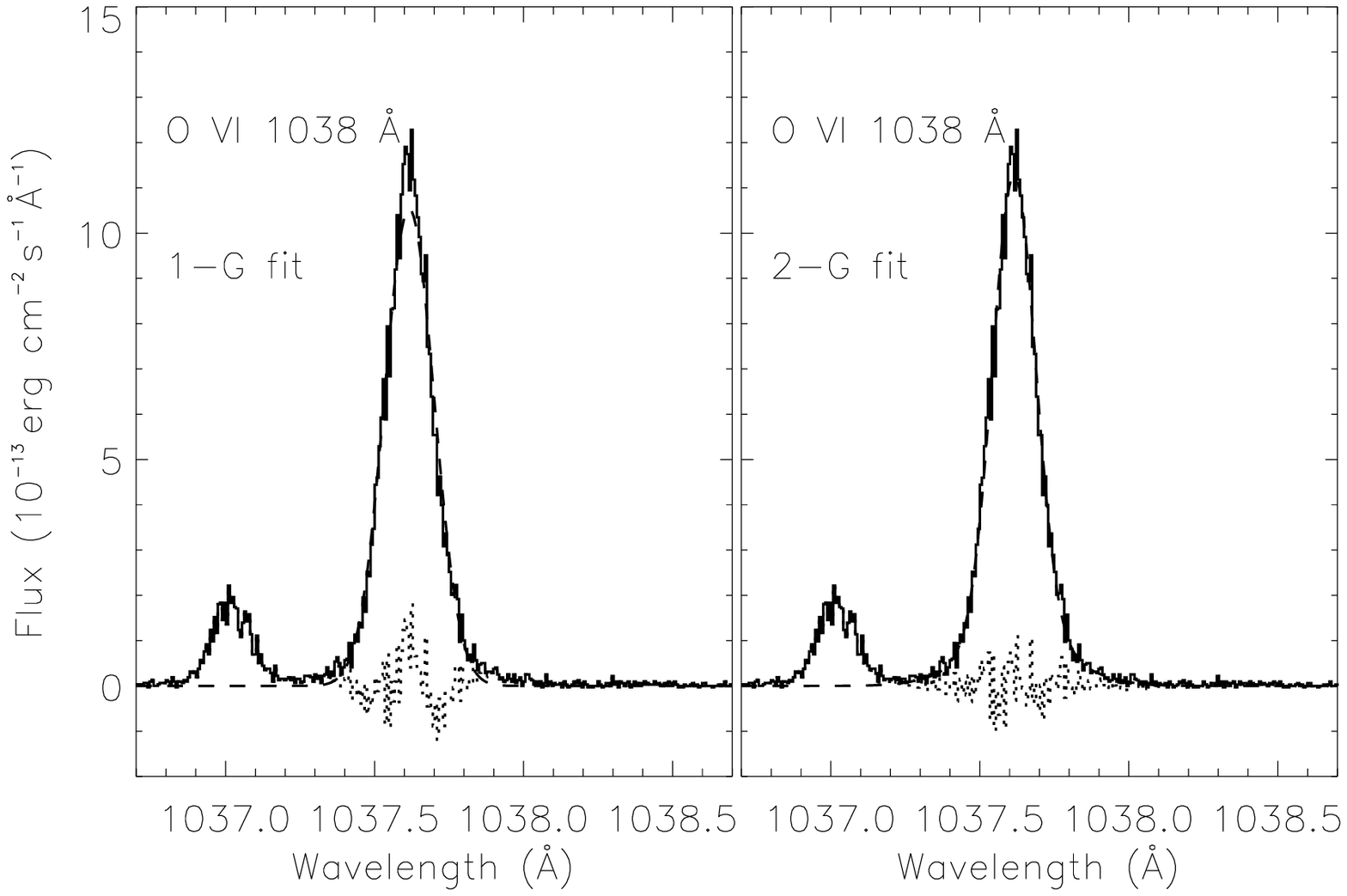, width=8.2cm, height=4.5cm}
\epsfig{file=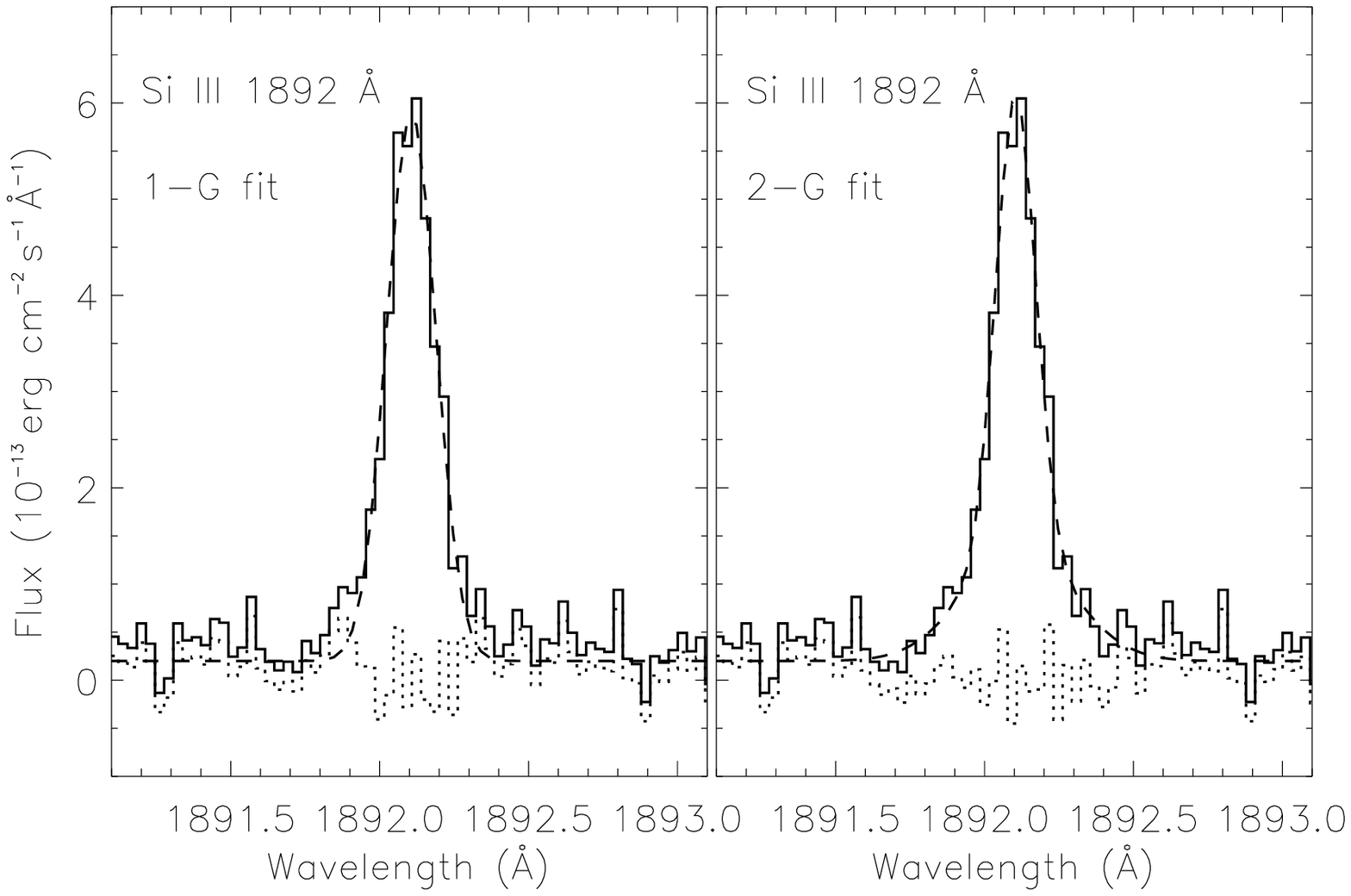, width=8.2cm, height=4.5cm}
\caption{Comparison of one-Gaussian fit (left) and two-Gaussian fit (right) for (from top left) the C\,{\sc iv} 1548-\AA~and 1550-\AA~lines, the Si\,{\sc iv} 1393-\AA~and 1402-\AA~lines, the N\,{\sc v} 1238-\AA~and 1242-\AA~lines, the O\,{\sc vi} 1032-\AA~and 1038-\AA~lines and the Si\,{\sc iii} 1892-\AA~line. The observed profiles are shown by the solid line, the fits by the dashed lines and the residuals by the dotted lines.}
\end{figure*}

It has previously been found (see Section 1) that the profiles of some 
transition region lines are described better by a two-Gaussian fit than a
one-Gaussian fit. This has been taken as an indication that there are (at
least) two separate atmospheric components contributing to the line
emission. This two-component nature of line widths has been studied in the 
Sun (Peter 2000a,b; 2001) and other stars (Wood et al. 1997). 
These authors have
found that line profiles for strong lines are described well by a sum
of two Gaussians, one narrow and one broad. For the Sun, Peter
(2000a,b; 2001)
found that the presence of the second, broad component is restricted
to the bright supergranulation network boundaries and that the 
cell interior regions show only the narrow component.

The excellent resolution and high signal-to-noise of the $\epsilon$
 Eri STIS spectra allow reliable two-component fits to be made to
 several of the strong spectral lines. The lines for which 
 two-component fits have been made (these were selected on the grounds that 
they are sufficiently strong to reliably identify a second component) are 
listed in Table~4, which also gives the flux
 ($F$), wavelength ($\lambda_{\mbox{\scriptsize obs }}$) and width
 ($\Delta \lambda$) for both the narrow and broad components. The
 error estimates quoted in Table~4 are formal 1$\sigma$ errors derived
 from the covariance matrix of the fit. The fluxes and widths are all
 strongly correlated while (with the exception of the N~{\sc v}~1242-\AA~
line) the wavelengths are only weakly correlated with the other
 parameters. Errors estimated this way do not explicitly include the
 limitation on measuring wavelengths due to the instrumental resolution,
 so the true uncertainties in $\lambda_{\mbox{\scriptsize obs}}$
 are larger than estimated here for the narrow components, and are similar to 
 the errors in $\lambda_{\mbox{\scriptsize obs}}$ as given in Table~1. 
Based on the line lists of Kurucz \& Bell (1995) and Sandlin et al. (1986),
no significant blends are expected in the wavelength ranges over which the 
fits to the profiles have been made.

For the strong lines, a two-component fit gives a significantly better 
description
of the data than does a one-component fit. To illustrate this, Fig.~11
shows the observed C\,{\sc iv}, Si\,{\sc iv}, N\,{\sc v}, O\,{\sc vi}
and Si\,{\sc iii}~(1892-\AA) lines with single-Gaussian fits and residuals 
(left-hand panels) and with double-Gaussian fits and residuals (right-hand 
panels). By
examining the residuals it is clear that the two-component fit gives
a substantial improvement for the
C\,{\sc iv} and Si\,{\sc iv} lines and some improvement for the other lines.
 To be more quantitative, the one- and two-Gaussian fits have
been used to calculate the reduced chi-squared ($\chi^{2}_{r}$)
statistic for each of the lines considered; these are given in Table~5 and show
 that the two-Gaussian fits give a significantly better description 
of the profiles than do the one-Gaussian fits. The reduction in 
$\chi^{2}_{r}$ is greatest for the C\,{\sc iv} and Si\,{\sc iv} lines but is 
still large enough to strongly suggest the presence of a second component in
 the N\,{\sc v}, O\,{\sc vi} and Si\,{\sc iii} lines. 

\begin{table}
\caption{Reduced chi-squared ($\chi^2_{r}$) for one- and two-Gaussian fits to selected strong lines.}
\begin{tabular}{llcc} \hline
Ion & $\lambda_{0}$ (\AA) & $\chi^2_{r}$ (one-Gaussian) & $\chi^2_{r}$ (two-Gaussian)\\ \hline
C {\sc iv} & 1548.204 & 14.28 & 0.87 \\
C {\sc iv} & 1550.781 & 8.24 & 1.40 \\
N {\sc v} & 1238.821 &  2.18 & 0.80 \\
N {\sc v} & 1242.804 & 2.20 & 1.22 \\
O {\sc vi} & 1031.912 & 5.25 & 1.34 \\
O {\sc vi} & 1037.613 & 3.11 & 1.18 \\
Si {\sc iii} & 1892.030 & 1.70 & 0.89 \\
Si {\sc iv} & 1393.760 & 18.04 & 2.18 \\
Si {\sc iv} & 1402.773 & 6.67 & 1.06 \\ \hline
\end{tabular}
\end{table}

As for the one-Gaussian fit parameters, the two-Gaussian fit
parameters (and associated 1$\sigma$ errors) have been converted to
velocity shifts ($v$) and most probable turbulent velocities ($\xi$)
for both components of each line. The results are given in Table 6. The 
temperatures at which both components 
are formed have been assumed to be the same as that used for the
one-Gaussian fits. The uncertainty in $\xi$ allows for a $\pm 0.1$~dex
uncertainty in $\log T_{\mbox{\scriptsize form}}$ in the 
same way as adopted for the one-Gaussian fits.
For O~{\sc vi} no velocity shift is given for the narrow component owing to
the uncertainty in the {\it FUSE} wavelength calibration. The velocity
shift given for the broad component is that relative to the narrow
component.

\begin{table*}
\caption{Line shifts ($v$) and turbulent velocities ($\xi$) for two-component fits. The subscripts N and B denote ``Narrow'' and ``Broad'' components respectively. The errors are deduced from the 1$\sigma$ errors in Table~4 and are given in same format as those in Table~2.}
\begin{tabular}{llllll}\\ \hline
Ion & $\lambda_{0}$ (\AA) & $v_{\mbox{\scriptsize N}}$ (km s$^{-1}$) & 
$\xi_{\mbox{\scriptsize N}}$ (km s$^{-1}$)&$v_{\mbox{\scriptsize B}}$ 
(km s$^{-1}$) & $\xi{\mbox{\scriptsize B}}$ (km s$^{-1}$) \\ \hline
C {\sc iv} & 1548.204 & $+3.3\pm0.2$  &
$15.91^{+1.23(0.69)}_{-1.52(0.92)}$ & $+8.3 \pm 0.8$ & 
$47.70^{+2.49(0.24)}_{-2.58(0.30)}$ \\[2mm]
C {\sc iv} & 1550.781 & $+0.6 \pm 0.4$  &
$13.27^{+1.80(0.82)}_{-2.29(1.11)}$
& $+4.6 \pm 0.8$ & $41.88^{+2.17(0.27)}_{-2.28(0.34)}$ \\[2mm]
N {\sc v} & 1238.821 & $+1.7 \pm 0.5$ &
$16.98^{+2.50(1.10)}_{-3.19(1.49)}$ 
&$+5.8 \pm 2.2$ & $53.05^{+7.36(0.36)}_{-7.64(0.46)}$ \\[2mm]
N {\sc v} & 1242.804 & $+3.2 \pm 0.7$ &
$16.13^{+3.09(1.15)}_{-4.04(1.58)}$ &
 $+14.7 \pm 6.8$ & $46.09^{+9.67(0.42)}_{-10.16(0.53)}$ \\[2mm]
O {\sc vi} & 1031.912 & see text & 
$18.82^{+2.30(1.53)}_{-3.09(2.12)}$ &
 $+5.5 \pm 1.7$ & $61.28^{+4.54(0.49)}_{-4.80(0.62)}$ \\[2mm]
O {\sc vi} & 1037.613 & see text & 
$17.70^{+2.66(1.62)}_{-3.65(2.28)}$ &
$+4.3 \pm 2.0$ & $63.86^{+6.87(0.47)}_{-7.19(0.59)}$ \\[2mm]
Si {\sc iii} & 1892.030 & $+11.9 \pm 0.6$ & $13.24^{+1.40(0.13)}_
{-1.50(0.16)}$ & $+13.0\pm3.5$ & $40.04^{+4.10(0.04)}_{-4.13(0.05)}$ \\[2mm]
Si {\sc iv} & 1393.760 & $+4.3 \pm 0.2$ &
$16.30^{+0.78(0.23)}_{-0.87(0.30)}$
& $+7.1\pm0.6$ & $48.57^{+1.64(0.08)}_{-1.67(0.10)}$ \\[2mm]
Si {\sc iv} & 1402.773 & $+4.5 \pm 0.4$ &$16.87^{+0.91(0.23)}_{-0.99(0.29)}$
& $+6.4 \pm 1.3$ & $54.47^{+3.43(0.07)}_{-3.46(0.09)}$ \\ \hline
\end{tabular}
\end{table*}

The red-shifts found for the narrow components are essentially the same as 
those found for the one-Gaussian fits.
A significant difference between the present results and the
solar results found by Peter (2000a,b; 2001) is that in $\epsilon$~Eri the 
broad component is {\it red-shifted} relative to the
narrow component, typically by about 4 km~s$^{-1}$. This can be seen directly
 in the asymmetry of the residuals shown in Fig.~11.  Peter (2000a) found that
 in observations of C\,{\sc iv} near Sun-centre the narrow component in the 
supergranulation cell boundaries is usually red-shifted, but the broad
component is red-shifted {\it less}. However, in Peter's (2000a)
detailed plots, there are a significant number of points where the
red-shift of the broad component exceeds that of the narrow component,
although these tend  to occur when the intensity of the broad
component is small. Since the relative intensity of the broad component in
$\epsilon$ Eri is larger than in the Sun (see below), this trend acts
in the wrong direction to account for our results.
In the cell interiors, where the broad component is absent,
the profiles show red-shifts comparable to those of the narrow components in
 the boundaries. In stellar observations of the C\,{\sc iv} lines
 Wood et al. (1997) observe a relatively greater red-shift of the broad 
component in the M-type flare star AU Mic, but this line was not observed in
the other less active main-sequence stars $\alpha$~Cen~A and B. However, in 
all three stars the Si\,{\sc iv} lines show relative shifts in the
same sense as in the Sun, in contrast to our results for $\epsilon$~Eri. 

\begin{figure}
\epsfig{file=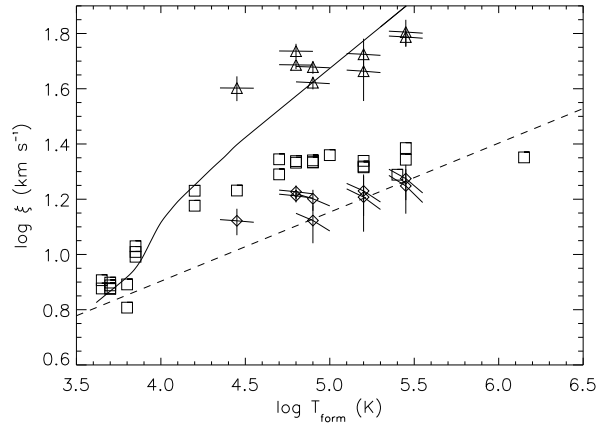, width=8.6cm}
\caption{$\log \xi$ versus $\log T_{\mbox{\scriptsize form}}$ for the 2-component fits (broad components shown by triangles; narrow components shown by diamonds). The single-component fits are shown by squares. The full line shows the sound speed and the dashed line shows a gradient of $\xi \propto T_{\mbox{\scriptsize form}}^{1/4}$, with an amplitude normalized to the O\,{\sc vi} narrow component. }
\end{figure}

Fig. 12 shows the non-thermal velocities of the two components as a
function of the line formation temperature. For comparison, the single
component points are shown again. The solid line shows the sound speed. The
dashed line shows a gradient corresponding to $\xi \propto T_{\mbox{\scriptsize
form}}^{1/4}$, normalized to pass through the O\,{\sc vi} narrow components.

The width of the broad component gives non-thermal velocities which
are greater than the sound speed for the Si~{\sc iii} and Si~{\sc iv}
lines, close to the sound speed for C~{\sc iv} and significantly
below the sound speed for N~{\sc v}. This behaviour is similar to that
found by Peter (2001) for the Sun and for the high gravity stars studied by
 Wood et al. (1997).  The widths of the low temperature lines from the 
single-Gaussian fits are also close to the sound speed (see Fig.~1 for the 
error bars).  

As in the Sun (Peter 2001), the ratio of the non-thermal widths of the two 
components is 
almost constant. The mean ratio in $\epsilon$~Eri is about
3, larger than the solar value (for lines in common) of about 2. Although 
the non-thermal widths of the broad components in $\epsilon$ 
Eri are on average similar to those in the Sun, the widths of the narrow 
component are smaller in $\epsilon$~Eri, by about a factor of 1.3.  Wood et
 al. (1997) found a systematic decrease in 
the width of the C\,{\sc iv} and Si\,{\sc iv} narrow components with 
increasing surface gravity. These widths in $\epsilon$~Eri fit their proposed
relation $\xi \propto g_{*}^{-0.68 \pm 0.07}$ remarkably well. A dependence
 on $g_{*}$, or alternatively the gas pressure, could indicate that turbulent
motions are ultimately involved in any wave generation. Peter (2001) 
suggests that the broad component
corresponds to a constant Alfv\'{e}n wave flux, with 
$\xi \propto T_{\mbox{\scriptsize form}}^{1/4}$. In $\epsilon$~Eri, 
to within the uncertainties, the same form would fit both components 
up to $T_{e} \simeq 3 \times 10^5$~K.
As stressed above, the temperature dependence expected for $\xi$
depends on the details of the wave periods and any departures from the WKB 
approximation.

Peter (2001) found that the relative fluxes in the narrow and
broad components vary with temperature, with the broad component having its 
largest
contribution at $\log T_{\mbox{\scriptsize form}} \approx 5.1$ with smaller
contributions at both lower and higher temperatures ($\sim$ 0.2 for
Si\,{\sc iv} and O\,{\sc vi} and $\sim$ 0.4 for C\,{\sc iv}). 
The ratio of the fluxes in the two components in  
$\epsilon$~Eri are given in Table~7. The 1$\sigma$ errors on the 
flux ratio are somewhat larger than might be expected given the errors in 
Table~4. This is
because of the strong anti-correlation between $F_{\mbox{\scriptsize B}}$ and 
$F_{\mbox{\scriptsize N}}$ which has been accounted for in
the 1$\sigma$ error estimates given in Table~7. In $\epsilon$~Eri the flux in
broad component is significantly larger, relative to the narrow one, 
than in the Sun. If the broad and narrow components genuinely represent
different physical regions, this suggests that the broad component may
increase with stellar activity. This was proposed by Wood et
al. (1997) on the basis of a larger sample of stars for which a
correlation with the stellar X-ray flux was apparent. 

\begin{table}
\caption{The ratio of the fluxes in the two-components including the 1$\sigma$ error estimates. The subscripts N and B denote ``Narrow'' and ``Broad'' components respectively.}
\begin{tabular}{llll}\\ \hline
Ion &  $\lambda_{0}$ (\AA) & $F_{\mbox{\scriptsize B}}/
F_{\mbox{\scriptsize N}}$ \\ \hline
C {\sc iv} & 1548.204 & $0.67 \pm 0.07$ \\
C {\sc iv} & 1550.781 & $1.04 \pm 0.15$ \\
N {\sc v} & 1238.821 & $0.45 \pm 0.13$\\
N {\sc v} & 1242.804 & $0.43 \pm 0.21$\\
O {\sc vi} & 1031.912 & $0.30 \pm 0.04$\\
O {\sc vi} & 1037.613 & $0.22 \pm 0.05$\\
Si {\sc iii} & 1892.030 & $0.66 \pm 0.13$\\
Si {\sc iv} & 1393.760 & $0.75 \pm 0.06$\\
Si {\sc iv} & 1402.773 & $0.61 \pm 0.07$ \\ \hline
\end{tabular}
\end{table}

\begin{figure}
\epsfig{file=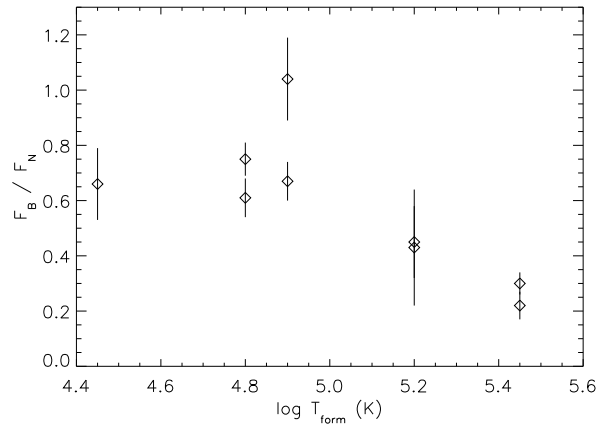, width=8.6cm}
\caption{Ratio of the flux emitted in the broad component to the that in the narrow component versus  $\log T_{\mbox{\scriptsize form}}$.}
\end{figure}

Fig. 13 shows a plot of the ratio of the line fluxes in the two
 components versus $\log T_{\mbox{\scriptsize form}}$. Although there
 is a significant difference between the flux ratio deduced from the two 
C\,{\sc iv} lines, the N\,{\sc v} and O\,{\sc vi} lines do show that the 
 broad component is less important at higher temperatures. In contrast
 to the results of Peter (2001, his fig. 6), in $\epsilon$~Eri  the 
Si\,{\sc iv} and Si\,{\sc iii} lines do not show a rapid decrease in the
 contribution of the broad component at lower temperatures. We point
 out that the sharp drop in the relative strength of the broad component at low
 temperatures deduced by Peter (2001) depends on his analysis of the 
Si\,{\sc ii} 1533-\AA~and C\,{\sc ii} 1335-\AA~lines (the latter is actually a
 close blend of two components of the same multiplet). Both of these
 lines have been excluded from the current study on the grounds that
 they have significant opacity and hence their profiles can be
 affected by radiative transfer effects. Also, models of the
 atmosphere (Sim 2002) suggest that the 
 temperatures of formation adopted by Peter (2001) for these lines 
 are too high, in which case the thermal
 broadening has been overestimated. The C\,{\sc ii} 1335-\AA~line in
 $\epsilon$~Eri does show broad wings and the analysis of this
 line profile will be discussed in later work on chromospheric
 modelling, including the radiative transfer.

Comparing Fig. 12 and 13 suggests a possible reason for the flattening off 
 of the non-thermal velocities with $T_e$ above $10^5$~K 
 when single Gaussian fits are used (see Fig. 1). Above this temperature
 the contribution of the broad component to the total line flux decreases
 so that the
 combined profile becomes dominated by the narrow component. Thus the
 single-Gaussian fit produces a smaller line width, even though the
 widths of both the narrow and broad component are not decreasing 
 with $\log T_{\mbox{\scriptsize form}}$. This explanation depends on the 
Fe {\sc xii} line flux being dominated by the narrow
 component, while the mid transition region lines Si~{\sc iii},
 Si~{\sc iv}, C~{\sc iv} and N~{\sc v} contain significant
 contributions from both components. 

\begin{figure}
\epsfig{file=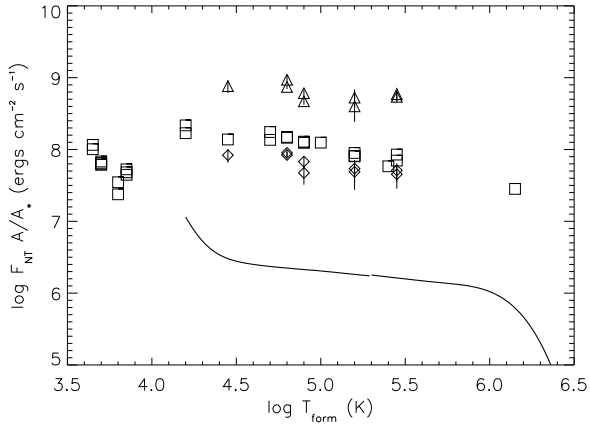, width=8.6cm}
\caption{$\log F_{\mbox{\scriptsize NT}} A/A_{*}$ versus $\log T_{\mbox{\scriptsize form}}$ for 2-component fits (broad components marked with a triangle and narrow components with a diamond). 
The fluxes are more constant than when a single-Gaussian fit to the line profiles is used. 
The boxes indicate results using
single-Gaussian fits, computed using $\rho_{ion}$. (See Fig. 5 for results
using $\rho$.)
The full line shows the flux required to account for the radiation losses above
a given $T_{e}$.}
\end{figure}

The non-thermal flux in each component for waves travelling at the
Alfv\'en speed, and assuming the same $P_e$ in both components, is
shown in Fig. 14.
Only calculations using $\rho_{ion}$ are shown since the 
lines for which two-Gaussian fits have been carried out
all form at sufficiently high temperatures
that $\rho_{ion} \approx \rho$.
The drop in the flux derived for the narrow component (spatially averaged 
as before) between $10^5$~K and 
$1.4 \times 10^6$~K is now only a factor of 1.8, compared with a factor of
4.5 when single-Gaussian fits are made. The fluxes are still 
large compared with those required and the wave propagation needs to be
investigated. We have no way of telling whether or not the flux in the broad 
component continues to higher temperatures. If it decreases to the
value derived from
 the Fe\,{\sc xii} line, then there are problems in disposing of the energy, 
unless $b$ and $P_e$ are small throughout the broad component region. 
There would be similar problems regarding deposition of the energy 
if the broad components were restricted to regions where the
magnetic flux does not continue to the corona.   
As 
stressed above the broad component flux could continue without being 
observable in the 
Fe\,{\sc xii} line profile. In this case, the correlation between the degree
of the contribution of the broad component to the line profiles and the 
X-ray flux, found by Wood et al. (1997) is surprising, since the radiative
losses should be dominated by the narrow component. It is possible that the 
X-ray flux and the broad components are related to different aspects of the 
magnetic field configuration, which themselves have systematic trends. Thus 
the origin of the broad component and its relation to coronal heating is 
still unclear. Energy balance models are required to explore this further.

At lower temperatures, extrapolating
 the results of Peter (2001) would suggest that the lines are
 dominated by the narrow component. Certainly, there is no strong
 evidence for the presence of a second component in the C~{\sc i} or
 O~{\sc i} intersystem lines. If these lines were dominated by the
 broad component, the large rise in
 $\xi$ (see Fig. 12) when combined with $\rho_{ion}$, would at face
 value imply an {\it
 increase} in the wave flux and an energy source in the low transition
 region/upper chromosphere. For example, Axford \& McKenzie (1992) have 
suggested that reconnection events involving small flux tubes at the level of
 the chromosphere or low transition region could generate Alfv\'en
 waves. These
could be high frequency waves which could propagate to the corona before 
being damped (McKenzie, Axford \& Banaszkiewicz 1995, McKenzie,
 Banaszkiewicz \& Axford 1997). If $\rho$ is adopted, the wave flux in the
 chromosphere exceeds that required at higher temperatures (see
 Section 3) and it is possible that the broadening of the chromospheric
 lines is the result of a different heating process and not a continuation of 
either transition region component. 
 
An alternative explanation of the line-widths (of one or both components) is 
that the ion and electron temperature are not equal and the widths reflect 
high ion temperatures (see Peter 2001). The idea is
attractive given observations of high O\,{\sc vi} ion temperatures in polar
coronal holes (see Cranmer et al. 1999 and references therein), and proposals
 by Marsch, Goertz \& Richter (1982) and
McKenzie et al. (1995, 1997) that damping of Alfv\'en waves by ion-cyclotron
 resonance could heat and drive the solar wind. 
However, to predict the resulting effects on ions in the transition region
requires detailed calculations.

\section{Conclusions}

The excellent
properties of the {\sc STIS} instrument have allowed us to determine
the non-thermal motions and line shifts in the outer atmosphere of
$\epsilon$~Eri through analyses of the profiles of optically thin
lines. The non-thermal velocity fields are important in semi-empirical 
modelling of the atmosphere and the results presented here have been used in 
such models (Sim 2002). 

The measured line widths have been used to derive the non-thermal
energy content of the atmosphere. The energy which could be carried in
acoustic or Alfv\'en waves has been investigated using the simple WKB 
approximation. The critical wave periods for this approximation to
hold have been expressed in terms of observable parameters, as far as
is possible. The unknown parameters are related to the magnetic field.  
With new models of the outer atmosphere more rigorous calculations could be 
made of how such waves may or may not propagate to the corona. 
As concluded from previous studies, the observations appear to rule out 
acoustic waves as the source of coronal heating. 

A detailed analysis of the two-component nature of the mid-transition
 region lines of Si\,{\sc iii}, Si\,{\sc iv}, C\,{\sc iv}, N\,{\sc v}
 and O\,{\sc vi}
 has been carried out, and the results compared with those found for
 the Sun and other main-sequence stars. It is found that the broad 
component contributes a
 larger fraction of the transition region line flux than in the Sun, 
supporting the suggestion by Wood et al. (1997)
 that this component to the flux increases with stellar activity. In
 agreement with analyses of solar data the
 broad component is found to be less important in the higher
 temperature lines of N\,{\sc v} and O\,{\sc vi}; it might not be significant 
in the Fe\,{\sc xii} line. If so, the correlation between broad transition 
region lines and stellar X-ray emission found by Wood et al. (1997) is hard to
 understand, unless both independently reflect systematic changes in the 
structure and strength of the magnetic fields.  In $\epsilon$~Eri, 
 the broad component is {\it red-shifted} relative to the narrow component, 
whereas in the Sun, relative blue shifts are found.

The narrow components are less broad than in the Sun and follow the 
correlation
between $\xi$ and $g_{*}$ found by Wood et al. (1997). We consider that 
there are fewer difficulties in relating the {\it narrow} component to
the passage of Alfv\'en 
waves which could heat the corona, in contrast to Peter's (2001) proposal
that this role is played by the broad component.  

A large amount of work remains to be done on the theoretical interpretation
of the observations. However, the quality of the results from {\sc STIS} 
places new constraints on proposed heating mechanisms.

\section*{Acknowledgments}

We wish to thank the referee, P. Ulmschneider, for several useful comments
on this manuscript.

SAS acknowledges financial support provided by a PPARC research 
studentship (PPA/S/S/1999/02862).

\section*{References}

\noindent
Achour H., Brekke P., Kjeldseth Moe O., Maltby P., 1995, \\
\indent ApJ, 453, 945 \\
Arnaud M., Rothenflug R., 1985, A\&AS, 60, 425\\
Axford W. I., McKenzie J. F., 1992, in Marsch E., Schwenn\\
\indent  R., eds, Solar Wind Seven. Pergamon, Oxford, p. 1 \\
Ayres T. R., Linsky J. L., Simon T., Jordan C., Brown \\
\indent  A., 1983a, ApJ, 274, 784 \\
Ayres T. R., Stencel R. E., Linsky J. L., Simon T., Jordan \\
\indent C., Brown A., Engvold O., 1983b, ApJ, 274, 801   \\ 
Berger R. A., Bruner E. C., Jr., Stevens R. J., 1970,  \\
\indent Sol. Phys., 12, 370 \\
Boland B. C., Engstrom S. F. T., Jones B. B., Wilson R. \\
\indent 1973, A\&A, 22, 161 \\
Boland B. C., Dyer E. P., Firth J. G., Gabriel A. H., \\
\indent Jones B. B., Jordan C., McWhirter R. W. P., Monk P., \\
\indent  Turner R. F., 1975, MNRAS, 171, 697  \\ 
Blair B. \& Andersson B.-G., 2001, The {\it FUSE} Observer's\\
\indent Guide, Version 3.0 \\
Brekke P., Hassler D. M., Wilhelm K., 1997, Sol. Phys.,\\
\indent 175, 349 \\
Brown A., Jordan C., Stencel R. E., Linsky J. L., Ayres \\
\indent T. R., 1984, ApJ, 283, 731   \\
Browning P. K., 1991, Plasma Phys. Control. Fusion, 33, \\
\indent 539 \\
Bruner E. C., McWhirter R. P. W., 1979, ApJ, 231, 551 \\ 
Bruner E. C., Jr., Jones R. A., Rense W. A., Thomas G. \\
\indent E., 1970, ApJ, 162, 28 \\
Carlsson M., 1986, Uppsala Astronomical Observatory, \\
\indent Report No. 33 \\
Chae J., Sch\"{u}hle U., Lemaire P., 1998, ApJ, 505, 957\\
Cook J. W., Cheng C.-C., Jacobs V. L., Antiochos S. K., \\
\indent  1989, ApJ, 338, 1176 \\ 
Cranmer S. R., Field G. B., Kohl J. L., 1999, ApJ, 518, 937 \\
Cross R., 1988, An Introduction to Alfv\'en Waves. Adam \\
\indent  Hilger, Bristol  \\
Dere K. P., Mason H. E., 1993, Sol. Phys., 144, 217 \\
Doschek G. A., Feldman U., VanHoosier M. E., Bartoe \\
\indent J.-D. F., 1976, ApJS, 31, 417    \\
Doyle J. G., O'Shea E., Erd\'elyi R., Dere K. P., Socker D. \\
\indent G., Keenan F. P. 1997, Sol. Phys., 173, 243 \\
Edl\'en B., 1934, Nova Acta R. Soc. Uppsala IV, 9, 6 \\
Feldman U., Doschek G. A., VanHoosier M. E., Purcell J. \\
\indent  D., 1976, ApJS, 31, 445 \\
Griesmann U., Kling R., 2000, ApJL, 536, L183 \\
Hallin R., 1966, Ark. Fys., 31, 511 \\
Hansteen V. H. 1993, ApJ, 402, 741 \\
Hollweg J. V., 1991, in Ulmschneider P., Priest E. R., \\
\indent Rosner R., eds, Mechanisms of Chromospheric and \\
\indent Coronal Heating. Springer-Verlag, Berlin, p. 423  \\
Jordan C., 1991, in Ulmschneider P., Priest E. R., Rosner \\
\indent R.,  eds, Mechanisms of Chromospheric and Coronal \\
\indent Heating. Springer-Verlag, Berlin, p. 300 \\
Jordan C., 2000, Plasma Phys. Control. Fusion, 42, 415 \\ 
Jordan C., McMurry A. D., Sim S. A., Arulvel M. 2001a, \\
\indent MNRAS, 322, L5 \\ 
Jordan C., Sim S. A., McMurry A. D., Arulvel M. 2001b, \\
\indent MNRAS, 326, 303 \\ 
Jordan C., Sim S. A., McMurry A. D. 2001c, in Conf. Proc. \\
\indent Cool Stars, Stellar Systems and the Sun. 12th \\
\indent Cambridge Workshop, in press\\
Jordan C., Ayres T. R., Brown A., Linsky J. L., Simon T., \\ 
\indent 1987, MNRAS, 225, 903 \\
Kelly R. L., 1987, J. Phys. Chem. Ref. Data, 16 (Suppl. 1), \\
\indent 1 \\
Kjeldseth-Moe O., Nicolas K. R., 1977, ApJ, 211, 579 \\
Kurucz R. L., Bell B. 1995, Atomic Line Data CD-ROM \\
\indent No. 23, Cambridge, Mass.: S.A.O. \\
Landi E., Landini M., Dere K. P., Young P. R., Mason \\
\indent H. E., 1999, A\&AS, 135, 339\\
Linsky J. L., Ayres T. R, 1978, 220, 619 \\
Linsky J. L., Wood B. E., 1994, ApJ, 430, 342 \\
Macpherson K. P., Jordan C., 1999, MNRAS, 308, 510 \\
Macpherson K. P., Jordan C., Smith G., 1999, in Vial J.-C.,\\
\indent Kaldeich-Sch{\" u}mann B., eds, Proc. 8th SOHO \\
\indent Workshop, ESA Special Publications 446, p. 461\\
McKenzie J. F., Axford W. I., Banaszkiewicz M., 1997, \\
\indent Geophys. Res. Lett., 24, 2877 \\
McKenzie J. F., Banaszkiewicz M., Axford W. I., 1995, \\
\indent A\&A, 303, L45 \\
Marsch E., Goertz C. K., Richter K., 1982, J. Geophys. \\
\indent Res., 5, 111 \\
Meyer F., 1976 in Bonnet R.-M., Delache Ph., eds, The \\
\indent Energy Balance and Hydrodynamics of the Solar \\
\indent Chromosphere and Corona. Nice Observatory, Nice, \\
\indent p. 111 \\
Montesinos, B., Jordan C., 1993, MNRAS, 264, 900\\
Narain, U., Ulmschneider, P., 1990, Space Sci. Rev., 54, 377\\
Narain, U., Ulmschneider, P., 1996, Space Sci. Rev., 75, 453\\
Osterbrock D. E., 1961, ApJ, 134, 347  \\
Peter H. 1999, ApJ, 516, 490 \\
Peter H. 2000a, A\&A, 360, 761 \\
Peter H. 2000b, A\&A, 364, 933 \\
Peter H. 2001, A\&A, 374, 1108\\
Peter H., Judge P. G. 1999, ApJ, 522, 1148 \\
R\"{u}edi L., Solanki S. K., Mathys G., Saar S. H. 1997, A\&A, \\
\indent 318, 429 \\
Saar S. H., Osten R. A. 1997, MNRAS, 284, 803\\
Sandlin G. D., Bartoe J.-D. F., Brueckner G. E., Tousey \\
\indent R., VanHoosier M. E., 1986, ApJS, 61, 801\\
Scharmer G. B. Carlsson M. 1985, J. Comp. Phys., 59, 56 \\
Seely J. F., Feldman U., Sch\"uhle U., Wilhelm K., Curdt \\
\indent W., Lemaire P., 1997, ApJL., 484, L87 \\
Sim S. A., 2002, D.Phil. Thesis, University of Oxford \\
Thatcher J. D., Robinson R. D., Rees D. E. 1991, MNRAS, \\
\indent 259, 14 \\
Vernazza J. E., Avrett E. H., Loeser R. 1981, ApJS, 45, 635 \\
Wood B. E., Linsky J. L., Ayres T. R. 1997, ApJ, 478, 745 \\
Zirker J.B. 1993, Sol. Phys., 148, 43 \\

\begin{appendix}
\section*{Appendix: Broad wings due to non-isotropic
velocity fields}
\renewcommand{\theequation}{A\arabic{equation}}
\renewcommand{\thetable}{A\arabic{table}}
\renewcommand{\thefigure}{A\arabic{figure}}

The purpose of this appendix is to draw attention to the possibility that
the broad wings in observed transition region line profiles (Section~5) 
might be due to a non-isotropic non-thermal velocity field,
rather than being the result of emission in two physically distinct 
atmospheric components.

In deriving equation (1), it is assumed that the non-thermal velocities
are described by a Maxwellian distribution in three-dimensions.
There is no direct observational evidence to justify this assumption and it is
interesting to examine the consequences of it being incorrect.

Suppose that the
non-thermal motions are the result of oscillations caused by transverse 
waves propagating along
magnetic field lines. Consider a small volume element of material which is 
threaded by magnetic field lines which are at an angle $\theta$
to the observers line-of-sight. The non-thermal motions
will be restricted to the plane perpendicular to the magnetic
field. Assuming, for simplicity, that the non-thermal motions
within this plane are described by a 2-D Maxwell distribution,
the observed profile of an emission line emitted by the material in the
volume element will be

\begin{equation} 
I(\lambda) = \frac{C}{\sqrt{b^2+a^2 \sin^2 \theta}}\exp( - \lambda^2/(b^2+a^2 \sin^2 \theta)) 
\end{equation}
where $\lambda$ is the wavelength measured relative to line centre,
$I(\lambda)$ is the intensity at wavelength $\lambda$ and $C$ is a constant.
$a$ and $b$ specify the widths of the 2-D Maxwell distribution of non-thermal
velocities and the 3-D Maxwell distribution of thermal velocities
respectively. Line broadening from sources other than thermal and
non-thermal motions have been neglected.

In order to obtain the profile that would be observed from the entire star
it is necessary to assume a distribution for the angle $\theta$.
Here we adopt the simplest assumption, 
that the 
distribution is spherically symmetric.
This would be a reasonable assumption for a star
if the magnetic field were
completely disordered (randomly oriented) on macroscopic scales. 
It would also be approximately valid if the field was purely radial
over an (unresolved) hemispherical stellar surface: the calculations
presented here are appropriate for emission from a hemisphere but they
do not account for the excess emission off the limb from material in the
back hemisphere of the star. If
emission from the back hemisphere were
included in the calculation it would enhance the strength of the broad
wings predicted in the 2-D case.
The spherically symmetric distribution 
will also be appropriate for spatially resolved observations of the
Sun, provided that the field is tangled on scales less than the
spatial resolution or contains both radial and 
closed loop field configurations.
If the field were locally ordered, then it would be necessary to adopt a more
complex distribution.
Adding up different $\theta$ angle strips from centre to limb, 
the total stellar profile in the presence of 2-D non-thermal
motions is

\[
I(\lambda)_{\mbox{\scriptsize 2-D}} \propto \int_{0}^{\pi/2} \frac{\sin \theta \; \mbox{d}\theta}{\sqrt{b^2+a^2 \sin^2 \theta}}\exp( - \lambda^2/(b^2+a^2 \sin^2 \theta)) \; .
\]
\begin{equation}
\end{equation}

Alternatively, consider the case of non-thermal motions generated by 
longitudinal waves propagating along the magnetic field. The
non-thermal velocities are then confined to only
one dimension, along the field.
By analogy with the case given above the total stellar profile is

\[
I(\lambda)_{\mbox{\scriptsize 1-D}} \propto \int_{0}^{\pi/2} \frac{\sin \theta \; \mbox{d}\theta}{\sqrt{b^2+a^2 \cos^2 \theta}}\exp( - \lambda^2/(b^2+a^2 \cos^2 \theta)) \; .
\]
\begin{equation}
\end{equation}
Figs.~A1 and A2 shows profiles calculated from equations~(A2)
and (A3), respectively. In both cases the thermal width
has been  chosen to be appropriate to the C~{\sc iv} lines, $b=0.05$~\AA.
The non-thermal width
parameter ($a$) was chosen arbitrarily (see below).
In each figure a single-Gaussian fit to the line and the residual for that 
fit are plotted. These should be compared with the fits and residuals for
the observed profiles shown in Fig.~11.
Two-Gaussian fits have been performed on the profiles calculated
from equations~(A2) and (A3). The results of these fits (fluxes and widths
for the two components) are given in Table~A1.

\begin{figure}
\epsfig{file=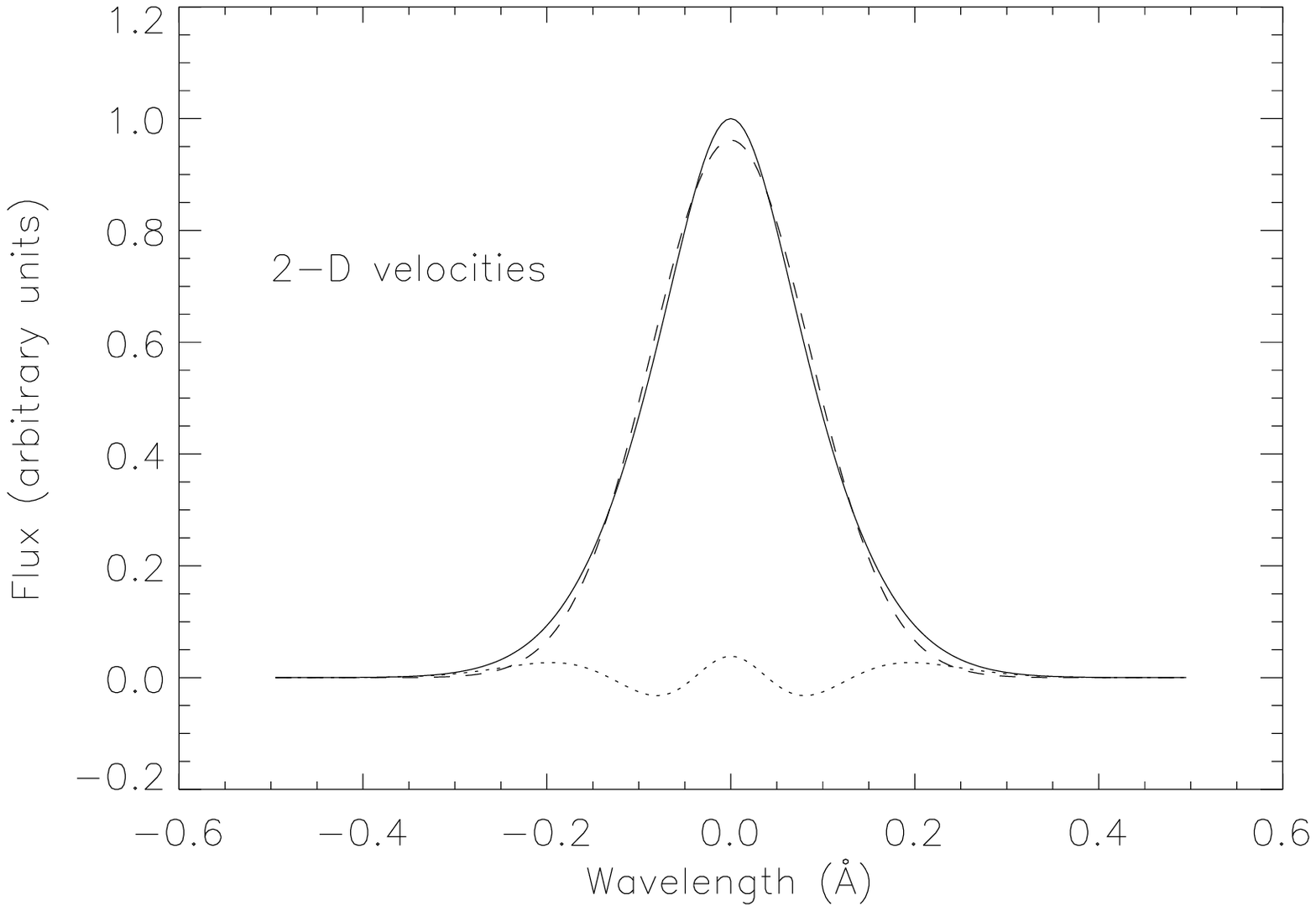, width=8cm, height=4cm}
\caption{Profile computed from Equation (A2) (solid line) with Gaussian over-plotted (broken line) and residual (dotted line). The values $a=0.15$~\AA~ and $b=0.05$~\AA~have been adopted.}
\end{figure}

\begin{figure}
\epsfig{file=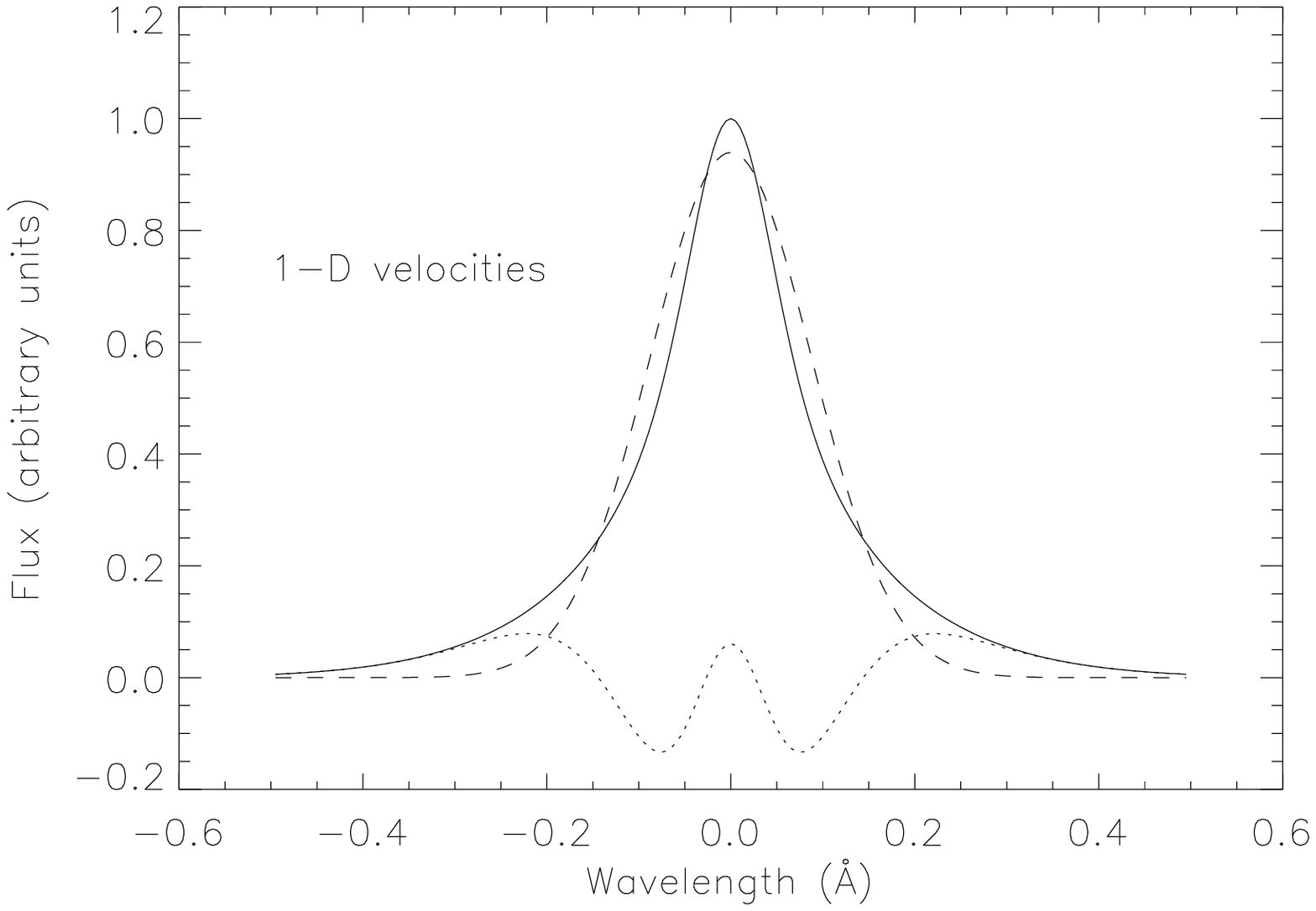, width=8cm, height=4cm}
\caption{Profile computed from Equation (A3) (solid line) with Gaussian over-plotted (broken line) and residual (dotted line). The values $a=0.32$~\AA~ and $b=0.05$~\AA~have been adopted.}
\end{figure}

\begin{table}
\caption{Two-Gaussian fit to the computed profiles for a 2-D non-thermal
velocity distribution (equation~A2) and a 1-D non-thermal velocity 
distribution (equation~A3). 
The fluxes ($F$) are given in arbitrary units and line widths 
($\Delta \lambda$, FWHM) are given in \AA.
The subscripts $N$ and $B$ refer to the narrow and broad Gaussian components.}
\begin{tabular}{l c c c c} \\ \hline
&$F_{N}$ & $\Delta \lambda_{N}$ & $F_{B}$ & $\Delta \lambda_{B}$  \\ \hline
2-D case&0.0571 & 0.137 & 0.1560 & 0.244 \\ 
1-D case& 0.0933 & 0.132 & 0.1253 & 0.394 \\ \hline
\end{tabular}
\end{table}

The values of $a$ used in the calculations were chosen so that the
width of the narrow Gaussian fit component
is  comparable to that observed for C~{\sc iv}. For the 2-D case
$a=0.15$~\AA~and for the 1-D case $a=0.32$~\AA.

Figs.~A1 and A2 show that these non-isotropic
velocity distributions produce line profiles that have broader wings than 
a Gaussian, in qualitative agreement with the observed profiles.
The broad wings are relatively stronger in the 1-D than 2-D case. 
The
width of the broad Gaussian component fitted to the calculated profile
is in very good agreement with that fitted to the observed C~{\sc iv}
profiles for the 1-D case, but is too narrow in the 2-D
case.
There is some quantitative disagreement between these calculations and the 
observations. In both cases, the flux ratio of the broad to narrow component
is too large, compared with the observations.
The 1-D flux ratio is consistent with
the observed 1550-\AA~flux ratio at the 2-$\sigma$ level, but is
very different from the 1548-\AA~ratio.
Also, the computed profiles are
symmetric about line centre while the observed broad components are
all redshifted relative to the narrow components by several km~s$^{-1}$.
In the model used here (which neglects the line formation process), to 
produce a redshifted broad component would require a mean net flow
away from the observer.
To shift the broad component relative to the narrow 
for the 1-D case would require a redshift that is largest in the regions
with small $\theta$, while for the 2-D case the redshift would need to be
largest in regions with large $\theta$.

In conclusion, it is apparent from Figs.~A1 and A2 that a non-isotropic
velocity distribution could predict line profiles with broad wings without the
need to invoke a physically distinct second component to the
emission. 
Although the discussion here has used the passage of transverse 
and longitudinal waves as examples, it is noted that the 
results are general to any situation where the non-thermal velocities
are confined to only one or two-dimensions, whatever their origin.
It is not proposed that the extremely simple one and two-dimensional
non-thermal velocity distributions used here are realistic; we wish only
to draw attention to the fact that such distributions can 
naturally lead to non-Gaussian profiles whose properties are 
similar to those of observed transition region lines.
\end{appendix}

\label{lastpage}
\end{document}